%% file: main.tex
\documentclass[%
aps,
prb,
reprint,
superscriptaddress,
amsmath,
amssymb,
]{revtex4-2}

\usepackage{graphicx}
\usepackage{diagbox}
\usepackage{xcolor}
\usepackage{amsmath,amssymb,amsthm,amsfonts}
\usepackage{bbm}
\usepackage{bm}
\usepackage[normalem]{ulem}
\input{preamble}
\begin{document}

\title{Projector Method for Nonlinear Light-Matter Interactions and Quantum Geometry}

\author{Zhichao Guo}
\thanks{These authors contributed equally to this work.}
\affiliation{Center for Quantum Matter, School of Physics, Zhejiang University, Zhejiang 310058, China}

\author{Zhuocheng Lu}
\thanks{These authors contributed equally to this work.}
\affiliation{Center for Quantum Matter, School of Physics, Zhejiang University, Zhejiang 310058, China}


\author{Hua Wang}
\email{daodaohw@zju.edu.cn}
\affiliation{Center for Quantum Matter, School of Physics, Zhejiang University, Zhejiang 310058, China}

\begin{abstract}
We develop a systematic projector-based Feynman diagram framework that intrinsically encodes quantum geometry for nonlinear optical responses. By explicitly incorporating geometric quantities such as the quantum geometric tensor, quantum hermitian connection, and triple phase product, the method ensures component-wise gauge invariance and seamlessly extends to multiband systems, enabling accurate calculations of quantum geometry and nonlinear optical responses. We derive the projector formalism in Wannier function basis and implement the \textit{ab initio} calculations of shift current in GeS, demonstrating excellent agreement with the sum rule and Wilson loop approaches. This work extends projector-based representations within the Wannier functions basis, offering an efficient and reliable tool for investigating nonlinear light-matter interactions and quantum geometry in realistic materials.
\end{abstract}

\pacs{}
\maketitle

\input{sections/section01.tex} 
\input{sections/section02.tex}

\input{sections/section03.tex}

\input{sections/section04.tex}
\input{sections/section05.tex}

\input{sections/acknowledgements.tex}

\bibliography{ref}


\clearpage
\onecolumngrid
\setcounter{section}{0}
\setcounter{figure}{0}
\setcounter{equation}{0}
\setcounter{table}{0}
\section*{Supplementary Material}
\input{SM/SM01}

\input{SM/SM02}

\input{SM/SM03}

\end{document}

%% file: preamble.tex
\usepackage{amsthm}
\usepackage{mathtools}
\usepackage{physics}
\usepackage{xcolor}
\usepackage{graphicx}
\usepackage{adjustbox}
\usepackage{placeins}
\usepackage[T1]{fontenc}
\usepackage{lipsum}
\usepackage{csquotes}
\usepackage{tikz}
\usepackage{tikz-feynhand} 
\usepackage{braket}
\usepackage{booktabs}
\usepackage{array} 
\usepackage{graphicx}
\usepackage{extarrows}
\usepackage{multirow}
\usepackage{float}

\usepackage[colorlinks,linkcolor=blue,anchorcolor=blue,citecolor=blue]{hyperref}

\usepackage{stackengine}

\usepackage{titlesec}
\titleformat{\paragraph}
  {\normalfont\normalsize\bfseries}{\theparagraph}{1em}{}

\titlespacing{\paragraph}{0pt}{5pt}{5pt} 

\usepackage{makecell}

%% file: sections/section01.tex
\section{Introduction}
   Quantum geometry has emerged as a fundamental concept in modern condensed matter physics, providing a powerful theoretical language for describing the nontrivial evolution of electronic states in parameter space. The concept first gained prominence with the discovery of the Berry curvature, a geometric quantity essential for understanding the integer quantum Hall effect\cite{Berry_Curvature_1980, Berry_Curvature_1995}. A complementary quantity, the quantum metric, was subsequently recognized\cite{QGT_Provost_1980, QGT_PATI_1991, QGT_Cheng_2013}, and is now understood to be crucial for phenomena 
  such as superconductivity\cite{SC_Paivi_2015, SC_Paivi_2017, QGT_SC_Bernevig_2020}, flat-band systems\cite{flatband_Kruchkov_2023, flatband_Huhtinen_2023}, and fractional Chern insulator\cite{FCI_Roy_2014, FCI_Mera_2021, FCI_Ledwith_2023}. Beyond the Berry curvature and quantum metric, recent years have witnessed the development of a more general theoretical framework for multiband quantum geometry\cite{Geometry_Ahn_PRX_2020, Geometry_Ahn_NP_2022}. In parallel, the concept of electronic geometry has been extended to interacting systems, including those with magnetoelectric \cite{magnetoelectric_2017, magnetoelectric_2021} and electron-phonon couplings\cite{EPC_Morimoto, EPC_Hu}. These advances naturally raise the question of how multiband quantum geometry can be effectively probed and manifested in physical observables.
    \par
    Nonlinear optical (NLO) phenomena have emerged as a powerful platform to probe multiband quantum geometry owing to their intrinsic multiband nature. In conventional NLO responses, the relevant geometric quantities typically involve two-band processes. For example, the injection current and shift current responses induced by monochromatic light have been linked to the interband quantum geometric tensor (interQGT) and the quantum Hermitian (Levi-Civita) connection (QHC), respectively\cite{Geometry_Ahn_PRX_2020}, while the photovoltaic Hall effect is closely associated with the Hermitian curvature\cite{Geometry_Ahn_NP_2022}. Our recent work has further advanced this direction by identifying a three-band quantum-geometric framework for photocurrents driven by bicircular light\cite{BCL_Guo_2025}. Given the strong connection between NLO responses and multiband quantum geometry, establishing a general theoretical framework linking the two would provide valuable guidance for probing quantum geometry through NLO responses.
    \par
    First-principles computational methods for NLO responses and quantum geometry have evolved significantly, with widely-used approaches including the traditional sum rule formalism \cite{Sum_rule_Julen_2018} and the generalized Wilson loop approach \cite{Generalized_Wilson_Wang2022}, providing an essential avenue for understanding a wide range of NLO phenomena. Recently, Avdoshkin et al. proposed a projector-based method for computing multi-state geometric quantities such as shift current and polarization\cite{proj_PRB, proj_PRL}, which has attracted considerable attention due to two key advantages: (i) the central object (projector) is manifestly gauge invariant \cite{projector_avdonshkin_2023}, whereas conventional methods often rely on gauge-dependent objects (e.g., Berry connections and generalized derivatives) that require additional treatments; and (ii) the projector formalism can, in principle, be extended to handle degeneracies \cite{projector_graf_2021, projector_pozo_2020}. However, studies using the projector formalism have primarily focused on simplified model systems and a limited set of optical responses. A major challenge remains in adapting this method to the Wannier function basis due to the intricate form of the matrix elements involved, and incorporating a broader range of light-matter interactions likewise also presents an open question.
    \input{Table/Feynman_diagram_components}
    \par
    In this work, we develop a systematic projector-based framework for identifying quantum geometry from NLO responses in realistic materials. Using a diagrammatic formulation, we derive the NLO conductivities and construct quantum geometry quantities within the projector formalism. As an example, we provide an explicit expression for the shift current and its associated geometric quantity. We then perform the first-principles calculations within the Wannier-based framework, addressing the challenges of band degeneracies with an effective scheme for accurate evaluation near these points. Finally, as a case study, we apply the projector-based method to calculate shift current in GeS and benchmark the results against those obtained from the conventional sum-rule method and the generalized Wilson loop approach to validate its accuracy.

%% file: Table/Feynman_diagram_components.tex
\usetikzlibrary{,fit}
\newcommand\addvmargin[1]{
  \node[fit=(current bounding box),inner ysep=#1,inner xsep=0]{};
}
\begin{table*}[t]
	\centering
	\small
	\begin{tabular}{c c c}
		\toprule
		\toprule
		Components & Diagram & Expression \\
		\midrule
        Input photon propagator &
        \begin{tikzpicture}[baseline=0] 
            \begin{feynhand}
                \vertex (a) at (0, 0) {};
                \vertex (b) at (2, 0) {};
                \propag[pho] (a) to [edge label = {$\omega_n$, $\alpha_n$}]  (b);
            \end{feynhand}
        \end{tikzpicture}
        & 1 
        \\     [0.4cm]
        Output current propagator &
        \begin{tikzpicture}[baseline=0] 
            \begin{feynhand}
                \vertex (a) at (0, 0) {};
                \vertex (b) at (2, 0) {};
                \propag[pho] (a) to [edge label = {$\omega_{\Sigma}$, $\mu$}]  (b);
            \end{feynhand}
        \end{tikzpicture}
        & 1 
        \\     [0.8cm]
        Electron propagator &
        \begin{tikzpicture}[baseline=0]
            \begin{feynhand}
                \vertex (a) at (0, 0) {};
                \vertex (b) at (2, 0) {};
                \propag[fer] (a) to [edge label = {$a$}]  (b);
            \end{feynhand}
        \end{tikzpicture}
        & $ G_a(\omega) = \left(\hbar\omega - \varepsilon_{a} + i\gamma\right)^{-1}$ 
        \\  [0.8cm]   
        Input vertex &
        \begin{tikzpicture}[baseline=0]
            \begin{feynhand}
                \vertex[dot] (a) at (0,0) {};
                \vertex (b) at (-0.71, 1.22) {$\omega_1$, $\alpha_1$};
                \vertex (c) at (-1.22, 0.71) {$\omega_2$, $\alpha_2$};
                \vertex (d) at (-1.0, -1.0) {$\omega_n$, $\alpha_n$};
                \propag[pho] (a) to (b);
                \propag[pho] (a) to (c);
                \propag[pho] (a) to (d);
                \vertex (e) at (-0.5, -0.2);
                \vertex (f) at (-0.6, 0.2);
                \propag[gho] (e) to [out = 120, in = -90] (f);
                \vertex (g) at (1, -1) {$a$};
                \vertex (h) at (1, 1) {$b$};
                \propag[fer] (g) to (a) ;
                \propag[fer] (a) to (h) ;
            \end{feynhand}
        \end{tikzpicture}
        & $ V^{\alpha_1 ...\alpha_n}_{ba} \equiv \frac{1}{n!}\prod\limits^{n}_{j=1} \left(\frac{ie}{\hbar \omega_j}\right)\hat{P}_b\partial^{(n)}_{\alpha_1 ...\alpha_n}\hat{H}\hat{P}_a$
        \\     [1.5cm]
        Output vertex &
        \begin{tikzpicture}[baseline=0]
            \begin{feynhand}
                \vertex[crossdot] (a) at (0,0) {};
                \vertex (b) at (+0.71, 1.22) {$\omega_{\Sigma}$, $\mu$};
                \vertex (c) at (+1.22, 0.71) {$\omega_1$, $\alpha_1$};
                \vertex (d) at (+1.0, -1.0) {$\omega_n$, $\alpha_n$};
                \propag[pho] (a) to (b);
                \propag[pho] (a) to (c);
                \propag[pho] (a) to (d);
                \vertex (e) at (+0.5, -0.2);
                \vertex (f) at (+0.6, 0.2);
                \propag[gho] (e) to [out = 60, in = -90] (f);
                \vertex (g) at (-1, 1) {$a$};
                \vertex (h) at (-1, -1) {$b$};
                \propag[fer] (g) to (a) ;
                \propag[fer] (a) to (h) ;
            \end{feynhand}
        \end{tikzpicture}
        & $ V^{\mu\alpha_1 ...\alpha_n}_{ba} \equiv  \frac{1}{n!}\frac{e}{\hbar}\prod\limits^{n}_{j=1} \left(\frac{ie}{\hbar \omega_j}\right) \hat{P}_b\partial^{(n+1)}_{\mu\alpha_1 ...\alpha_n}\hat{H}\hat{P}_a $
        \\ [1.5cm]
		\bottomrule
		\bottomrule
	\end{tabular}
	\setlength{\abovecaptionskip}{0.4cm}
	\caption{Diagram components and Feynman rules for NLO responses. The input vertex includes only the contributions from input fields $\{\omega_j, \alpha_j\}$, while the output vertex may contain both the output current $\{\omega_{\Sigma}, \mu\}$ and the contributions from input fields  $\{\omega_j, \alpha_j\}$. Here we denote $\omega_{\Sigma} = \sum^{\text{All Input}}_{n=1} \omega_{n}$ as the total frequency of all input light fields. Here, $a$ and $b$ label subspaces with energy $\varepsilon_a$ and $\varepsilon_b$, respectively.}
	\label{feynman rule}
\end{table*}

%% file: sections/section02.tex
\section{Feynman Diagram in Projector Formalism}
    \label{Projector Formalism for Nonlinear Optics}
    In this section, we outline a general procedure for extracting the quantum-geometric components of NLO responses within the projector formalism, starting from the diagrammatic method\cite{Diagrammatic_Parker_Moore_2019}. The computational procedures are subsequently illustrated using shift current as a representative example.
    
    \subsection{Generalities}
        For a crystalline material, the single-particle Hamiltonian in second quantization is expressed as:
        \begin{equation}
            \hat{H}_0 = \sum\limits_{a} \int [d\boldsymbol{k}] \varepsilon_{a}(\boldsymbol{k})\hat{P}_{a}(\boldsymbol{k}),
        \end{equation}
        $[d\boldsymbol{k}]$ is defined as intergral over the Brillouin zone $d^{d}\boldsymbol{k}/(2\pi)^d$, with $d$ being the dimension of the system. The summation $\sum_a$ is taken over all subspaces, each containing eigenstates with the same energy. For a subspace containing $N_s$ eigenstates ${\ket{u_{a_i}}}$, the corresponding band energy is $\varepsilon_a$, and its projector is given by $\hat{P}_a = \sum_{i=1}^{N_s} \ket{u_{a_i}}\bra{u_{a_i}}$. This formulation applies to nondegenerate cases and, more importantly, naturally generalizes to degenerate cases, ensuring that all subsequent derivations remain valid in the presence of band degeneracies.
        \par
        According to the diagrammatic method, the optical response of such a system can be evaluated following the Feynman rules outlined below:
        \par
        (1) The conductivity is formulated as a sum over closed-loop Feynman diagrams. The analytic expressions corresponding to the components of each diagram are summarized in Table~\ref{feynman rule}. Each diagram contains one output vertex and a number of input vertices, possibly zero. In particular, when evaluating the \textit{n}th-order optical response, the relevant diagrams must satisfy the following criteria: they contain all possible configurations with $n$ input photon propagators, each labeled by a Cartesian index $\alpha_1 \ldots \alpha_n$, and one output current propagator labeled by a Cartesian index $\mu$. The frequency $\omega_{\Sigma}$ associated with the output current is given by the sum of the input light frequencies, in accordance with energy conservation.
        \par
        (2) The \textit{n}-th conductivity $\sigma$ consists of two components: the vertex product (V) and the Green's function product (G):
            \begin{equation}
                \begin{aligned}
                    \sigma
                    =& \sum \Trace{\left(\prod\limits^{\text{All Vertices}}_{i=1} V_i\right)} \times \left(\prod\limits^{\text{All Inner Lines}}_{j=1} G_j\right) \\ =& \sum V \times G,
                \end{aligned}
            \end{equation}
        where $V_i$ and $G_j$ denote the vertices and Green's functions (electron propagators), respectively, as summarized in Table~\ref{feynman rule}. For clarity, Cartesian and subspace indices are suppressed. The full conductivity is obtained by summing over subspace indices in $V$ and $G$ and integrating over the Brillouin zone, as indicated by the $\sum$ symbol. The vertex product given by the trace over all vertices is of particular interest, as it encodes essential geometric information. In contrast, the Green’s function product, evaluated via Matsubara summation, captures the joint density of states and governs the resonance structure of the optical responses.
        \par
        The vertex product can be systematically expanded according to a set of rules, yielding expressions in terms of energies, projectors, and their derivatives. Below we list the expansion rules for first- and second-order derivatives of the Hamiltonian, which are most commonly used:
        \begin{equation}
            \begin{aligned}
                \hat{P}_a \left(\partial_{\alpha}\hat{H}\right) \hat{P}_b = \delta_{\varepsilon_{ab}} \left(\partial_{\alpha}\varepsilon_{a}\right)\hat{P_a} + (1-\delta_{\varepsilon_{ab}})\varepsilon_{ba}\hat{P}_{a}\hat{P}^{\alpha}_{b},
            \end{aligned}
        \end{equation}
        \begin{equation}
            \begin{aligned}
                &\hat{P}_a\left(\partial_{\alpha}\partial_{\beta}\hat{H}\right)\hat{P}_b \\
                =&\delta_{\varepsilon_{ab}}\left(\partial_{\alpha}\partial_{\beta}\varepsilon_{b} \right)\hat{P}_a + \varepsilon_{ba}\hat{P}_a\left(\hat{P}^{\alpha\beta}_b + \hat{P}^{\alpha}_a\hat{P}^{\beta}_b\right) \\
                &+  \left(\partial_{\alpha}\varepsilon_{ba}\right)\hat{P}_a\hat{P}^{\beta}_{b} + \left(\partial_{\beta}\varepsilon_{ba}\right)\hat{P}_a\hat{P}^{\alpha}_{b}\\
                &+ \sum\limits_{\varepsilon_{c}\neq \varepsilon_{a}; \varepsilon_{c}\neq \varepsilon_{b}}  \left(\varepsilon_{ac}\hat{P}_a\hat{P}^{\beta}_{c}\hat{P}^{\alpha}_{b}- \varepsilon_{cb}\hat{P}_a\hat{P}^{\alpha}_{c}\hat{P}^{\beta}_{b}\right).
            \end{aligned}
        \end{equation}
        We use the shorthand notation $\partial_{\alpha} \equiv \partial/\partial k_{\alpha}$ for the partial derivative with respect to $k_{\alpha}$; $\varepsilon_{ab} = \varepsilon_a - \varepsilon_b$ for the energy difference between subspaces $a$ and $b$, with the explicit $\boldsymbol{k}$ dependence omitted for clarity; $\delta_{\varepsilon_{ab}}$ as a Kronecker delta defined by $\delta_{\varepsilon_{ab}}=1$ when $\varepsilon_{ab}=0$ and $\delta_{\varepsilon_{ab}}=0$ otherwise; and $\partial_{\alpha}\hat{P} \equiv \hat{P}^{\alpha}$, $\partial_{\alpha}\partial_{\beta}\hat{P} \equiv \hat{P}^{\alpha\beta}$ for projector derivatives.
        \par
        With the aid of these expansion rules, the vertex product can be recast in the projector formalism. This framework also enables the systematic identification of all geometric quantities, thereby facilitating for the implementation of numerical algorithms. As an illustration, we perform the first-principles calculation of the shift current conductivity within the projector formalism.
        
    \subsection{Shift Current in Projector Formalism}
        \label{sec:Shift Current in Projector Formalism}
        \input{Figure/feynman_2nd_combine}
        The shift current is a manifestation of the bulk photovoltaic effect (BPVE) and can occur in noncentrosymmetric homogeneous materials \cite{BPVE_Fridkin_2001, BPVE_Wolfgang_1981, BPVE_Sipe_2000, BPVE_Andrew_2012, BPVE_Tan_2016, BPVE_Morimoto_2016}, in contrast to the conventional photovoltaic effect that requires a $p$–$n$ junction. Notably, it is a resonant response that generates photocurrent instantaneously, independent of relaxation time. The Feynman diagrams corresponding to the shift current are shown in Fig. \ref{2nd feyn diagram}, involving the first- and second-order derivatives of the Hamiltonian. The vertex products for the two-vertex diagram, $V^{\mu\alpha\beta, \text{sh(2)}}_{ab}$, and the three-vertex diagram, $V^{\mu\alpha\beta, \text{sh(3)}}_{abc}$, are given by:
        \begin{equation}
            V^{\mu\alpha\beta, \text{sh(2)}}_{ab} = \frac{e^3}{\hbar^3\omega^2}\text{Tr}\left[\hat{P}_b\left(\partial_{\alpha}\hat{H}\right)\hat{P}_{a}\left(\partial_{\mu}\partial_{\beta}\hat{H}\right)\hat{P}_b\right],
        \end{equation}
        \begin{equation}
            V^{\mu\alpha\beta, \text{sh(3)}}_{abc} = \frac{e^3}{\hbar^3\omega^2}\text{Tr}\left[\hat{P}_b\left(\partial_{\alpha}\hat{H}\right)\hat{P}_{a}\left(\partial_{\mu}\hat{H}\right)\hat{P}_{c}\left(\partial_{\beta}\hat{H}\right)\hat{P}_b\right].
        \end{equation}
        We then apply the expansion rules to extract the quantum geometric quantities from the vertex products within the projector formalism. By attaching the Green’s function products and combining the contributions from both diagrams, we obtain a fully gauge-invariant expression for the shift-current conductivity, with the detailed derivation provided in Sec. I of the Supplemental Material:
        \begin{equation}
            \begin{aligned}
                &\sigma^{\mu\alpha\beta}_{\text{sh}} \\
                =& -\frac{i\pi e^3}{\hbar^4\omega^2}\int[d\boldsymbol{k}]\sum\limits_{a, b}\varepsilon^2_{ab}\text{Tr}\left[\hat{P}_b\hat{P}^{\alpha}_a\left(\hat{P}^{\mu\beta}_b + \hat{P}^{\mu}_a\hat{P}^{\beta}_b\right)\right] \\
                &\times f_{ab}\delta(\hbar\omega +\varepsilon_{ab})+ \left[(\alpha, \beta, \omega)\leftrightarrow(\beta, \alpha, -\omega)\right] \\
                =& -\frac{i\pi e^3}{\hbar^2}\int[d\boldsymbol{k}]\sum\limits_{a, b}\left(C^{\mu; \alpha\beta}_{ab} - C^{\mu; \beta\alpha}_{ba}\right)f_{ab}\delta(\hbar\omega +\varepsilon_{ab}).\\
            \end{aligned}
        \end{equation}
        Here, $e = -|e|$ is the electron charge, and $f_{ab}$ is the difference between the Fermi–Dirac occupations of subspace $a$ and $b$. The quantity $C^{\mu; \alpha\beta}_{ab} = \text{Tr}\left[\hat{P}_b \hat{P}^{\alpha}_a\left(\hat{P}^{\mu\beta}_b + \hat{P}^{\mu}_a \hat{P}^{\beta}_b\right)\right]$ is the QHC \cite{ Geometry_Ahn_NP_2022, proj_PRL}. It's a third order geometric tensor defined by the combined order of derivatives. 
        For a time-reversal-symmetric material irradiated by linearly polarized light (LPL) \cite{CPL_Dai_2023}, symmetry analysis reveals that $\text{Im}\left(C^{\mu; (\alpha\beta)}_{[ab]}\right) = -2iR^{\alpha; \mu}_{ab}r^{\alpha}_{ab}r^{\beta}_{ba}$ is the only surviving geometry part, where $(\alpha\beta)$ and $[ab]$ denote symmetrization over $\alpha,\beta$ and antisymmetrization over $a,b$, respectively. The quantity $r^{\alpha}_{ab} = \braket{a|i\partial_{\alpha}b}$ is the Berry connection, while $R^{\alpha; \mu}_{ab} = -\partial_{\mu}\phi^{\alpha}_{ab} + r^{\mu}_{aa} - r^{\mu}_{bb} $ is the well-known shift vector, with $\phi^{\alpha}_{ab}$ the phase of $r^{\alpha}_{ab}$. Physically, the shift vector represents the displacement of the electronic wave packet induced by optical excitation \cite{Physical_illustration_of_shift_vector, shiftvector_lu2025}.

%% file: Figure/feynman_2nd_combine.tex
\begin{figure}[ht]
    \centering
    \includegraphics[scale=0.6]{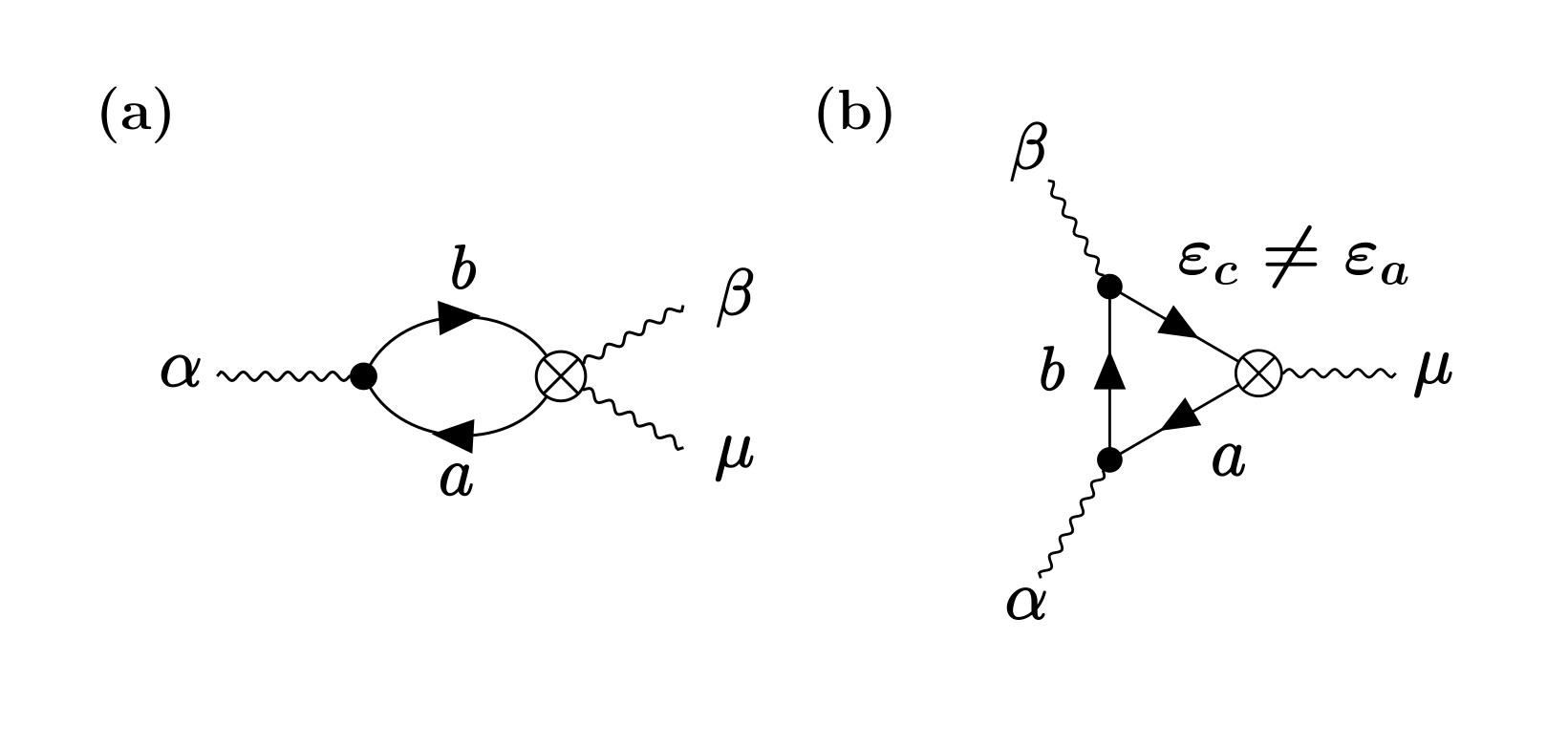}
    \caption{Feynman diagrams corresponding to the shift current. (a) and (b) are two-vertices and three-vertices diagram, respectively. The indices ${\alpha, \beta}$ and ${\mu}$ denote the Cartesian components of the three incident electric fields and the output current, respectively, while ${a, b, c}$ label the band indices of each Green’s function. In diagram (b), the shift-current contribution arises only from processes involving bands $a$ and $c$ with different energies, i.e., $\varepsilon_c \neq \varepsilon_a$.}
    \label{2nd feyn diagram}
\end{figure}

%% file: sections/section03.tex
\section{Wannier-Based Projector Approach to Quantum Geometry}
    This section details the implementation of projector-based geometric quantities within first-principles calculations using Wannier interpolation. We also address the numerical challenges posed by band degeneracies and present practical strategies for their effective treatment.
    
    \subsection{Wannier representation}
    We begin by reviewing the construction scheme for Wannier functions (WFs) \cite{WFs_Vanderbilt, WFs_Wang}. While direct first-principles calculations of optical responses are generally computationally demanding, building a tight-binding (TB) model from WFs provides an efficient and accurate alternative. This approach can closely reproduce the Hamiltonian and band structure from first-principles calculations, ensuring that the computed optical and transport properties remain in good agreement.
    \par
    To construct a physically meaningful set of WFs, we first identify the atomic orbitals that contribute most significantly to the bands near the Fermi level. This analysis determines the energy window and the number of WFs, $N_{\text{wan}}$, required to span the relevant subspace. A standard disentanglement procedure is then performed, yielding a set of $N_{\text{wan}}$ disentangled WFs, $\ket{i, \boldsymbol{R}}$, per unit cell. Based on these WFs, a TB model can be constructed as follows:
    \begin{equation}
        \hat{H}(\boldsymbol{k}) = \sum\limits^{N_\text{{wan}}}_{i, j}\ket{i, \boldsymbol{k}} H_{ij}(\boldsymbol{k}) \bra{j, \boldsymbol{k}},
    \end{equation}
    where $\ket{i, \boldsymbol{k}} = \sum_{\boldsymbol{R}}e^{i\boldsymbol{k}\boldsymbol{R}}\ket{i, \boldsymbol{R}}$ are Wannier-gauge Bloch basis formed by Fourier transformation of Wannier functions. It is important to note, however, that disentangled WFs do not form an “exact TB basis” in the sense of being perfectly localized; rather, they retain a finite spatial spread. This spread plays a crucial role in the evaluation of projectors and their derivatives, as discussed in Sec.~\ref{Projector in Wannier Basis}.
    \par
    After constructing the TB model from WFs, we diagonalize the Hamiltonian in the Wannier representation to obtain the band energies and eigenstates:
    \begin{equation}
        \hat{H}(\boldsymbol{k}) \hat{\mathcal{U}}(\boldsymbol{k}) = \hat{\mathcal{U}}(\boldsymbol{k}) \hat{E}(\boldsymbol{k}).
    \end{equation}
    Here, $\hat{E}(\boldsymbol{k})_{ab} = \delta_{ab}\varepsilon_{a}(\boldsymbol{k})$ is the diagonal band-energy matrix, and $\hat{\mathcal{U}}(\boldsymbol{k}) = \big[\ket{u_{1}(\boldsymbol{k})}, \dots, \ket{u_{N_{\text{wan}}}(\boldsymbol{k})}\big]$ collects the eigenstates, with $\ket{u_{a}(\boldsymbol{k})}$ denoting the $a$th eigenstate at momentum $\boldsymbol{k}$. Each eigenstate can be expanded in the Wannier-gauge Bloch basis as $\ket{u_{a}(\boldsymbol{k})} = \sum_{i} \ket{i, \boldsymbol{k}} U^{\dagger}_{ia}(\boldsymbol{k})$, where $U^{\dagger}_{ia}(\boldsymbol{k}) = \braket{i,\boldsymbol{k}|u_{a}(\boldsymbol{k})}$ serves as the expansion coefficient.
    \par
    
    \input{Table/Geometry_quantities_inter}
    \subsection{Projector Formalism in Wannier Representation}
    \label{Projector in Wannier Basis}
    Once the eigenvalues and eigenvectors of the Wannier TB model are obtained, the projectors and their derivatives in the Wannier representation can be systematically constructed. For clarity, we denote the matrix of an operator $\hat{\mathcal{O}}$ in the Wannier representation by $\hat{\mathcal{O}}^{(\text{W})}$, and in the Hamiltonian gauge by $\hat{\mathcal{O}}^{(\text{H})}$, which can be directly computed from the eigenstate matrix. We begin by examining the computation of the projector $P^{(\text{W})}_{a}$ in the Wannier representation:
    \begin{equation}
        \begin{aligned}
            \left(\hat{P}^{(\text{W})}_{a}\right)_{jl} =& \braket{j|u_a}\braket{u_a|l} = \sum\limits_{mn}\braket{j|m}U^{\dagger}_{ma}U_{an}\braket{n|l} \\
            =&  U^{\dagger}_{ja}U_{al} \equiv \left(\hat{P}^{(\text{H})}_{a}\right)_{jl}.
        \end{aligned}
    \end{equation}
    From this, we see that the projector $\hat{P}^{(\text{H})}_{a}$ in Hamiltonian gauge is equal to the desired projector $\hat{P}^{(\text{W})}_{a}$ in the Wannier representation.
    \par
    However, unlike the projector itself, the finite spread of the Wannier functions introduces non-negligible effects in the calculation of projector derivatives. Consequently, the derivatives of the projector cannot be directly replaced by their counterpart in Hamiltonian gauge, as this would lead to deviations from the true values. In the following, we derive the expressions for the projector derivatives $\partial_{\alpha}\hat{P}^{(\text{W})}_{a}$ and $\partial_{\alpha}\partial_{\beta}\hat{P}^{(\text{W})}_{a}$, fully taking into account the spread of the Wannier basis. The matrix elements of the first-order derivative $\hat{P}^{\alpha (\text{W})} \equiv \partial_{\alpha}\hat{P}^{(\text{W})}$ is given by:
    \begin{equation}
        \begin{aligned}
            \label{wannier 1st diff}
            \left(\hat{P}^{\alpha (\text{W})}_{a}\right)_{jl} =& \braket{j|\hat{P}^{\alpha}_{a}|l} \\
            =& \left(\hat{P}^{\alpha(\text{H})}_{a} - i A^{\alpha}\hat{P}^{(\text{H})}_{a} + i\hat{P}^{(\text{H})}_{a}A^{\alpha}\right)_{jl} \\
            \neq & \left(\hat{P}^{\alpha(\text{H})}_a\right)_{jl},
        \end{aligned}
    \end{equation}
    where $A^{\alpha}(\boldsymbol{k})$ is the Fourier transform of the WFs dipole matrix, i.e. $A^{\alpha}_{jl}(\boldsymbol{k}) = \braket{j, \boldsymbol{k}|i\partial_{\alpha} l, \boldsymbol{k}} = \sum_{\boldsymbol{R}}e^{i\boldsymbol{k}\boldsymbol{R}}\braket{j, \boldsymbol{0}|\hat{r}_{\alpha}|l, \boldsymbol{R}}$. The quantity $\partial_{\alpha}\hat{P}^{(\text{H})}_a \equiv \hat{P}^{\alpha(\text{H})}_a$ denotes the first-order derivative of the projector in Hamiltonian gauge, which can be computed using finite differences due to its gauge-invariant nature.
    \par
    Similarly, the second order derivative of the projector in Wannier representation can be written as
    \begin{equation}
        \begin{aligned}
            \label{wannier 2nd diff}
            &\left(\hat{P}^{\alpha\beta(\text{W})}_{a}\right)_{jl} = \braket{j|\hat{P}^{\alpha\beta}_{a}|l}
            \\
            =& \left(\frac{1}{2}\hat{P}^{\alpha\beta(\text{H})}_a + i\hat{P}^{\beta(\text{H})}_a A^{\alpha}-iA^{\alpha}\hat{P}^{\beta(\text{H})}_a + A^{\alpha}\hat{P}^{(\text{H})}_a A^{\beta}\right)_{jl} \\
            & - \left(\frac{1}{2}D^{\dagger, \alpha\beta} \hat{P}^{(\text{H})}_a+ \frac{1}{2}\hat{P}^{(\text{H})}_a D^{\alpha\beta} \right)_{jl}+ (\alpha \leftrightarrow \beta)\\
             &\neq \left(\hat{P}^{\alpha\beta(\text{H})}_a\right)_{jl},
        \end{aligned}
    \end{equation}
    where $D^{\alpha\beta}(\boldsymbol{k})$ is the Fourier transform of the WFs quadrupole matrix, defined as $D^{\alpha\beta}_{jl}(\boldsymbol{k}) = -\braket{\partial_{\alpha}\partial_{\beta}j, \boldsymbol{k}|l, \boldsymbol{k}} = \sum_{\boldsymbol{R}}e^{i\boldsymbol{k}\boldsymbol{R}}\braket{j, \boldsymbol{0}|\hat{r}_{\alpha}\hat{r}_{\beta}|l, \boldsymbol{R}}$. $D^{\dagger, \alpha\beta}$ is defined as the complex conjugate of $D^{\alpha\beta}$. The second-order derivative of the projector in the Hamiltonian gauge, $\partial_{\alpha}\partial_{\beta}\hat{P}^{(\text{H})}_a \equiv \hat{P}^{\alpha\beta(\text{H})}_a$, can be computed using finite differences in the same manner as the first-order derivative. We conclude that directly applying the projector formalism to the Wannier TB model is insufficient unless the derivatives of the WFs are properly incorporated. The computational strategy presented here can be easily generalized to other projector-related geometric quantities in the Wannier representation.
    \par
    Next, using the projectors and their derivatives derived above in the Wannier representation, we recast commonly used interband geometric quantities, such as the interQGT, QHC, and triple phase product (TPP), into forms suitable for numerical evaluation using Wannier interpolation. The general projector expressions of geometric quantities in Wannier representation and related physical phenomena are summarized in Table~\ref{geometry quantities}. This reformulation follows the guiding principles aimed at computational efficiency: Insert projectors on both sides of projector derivatives whenever possible, without approximation. When full decomposition is not feasible, we ensure that each Wannier-interpolated matrix contains derivatives of the lowest possible order, as higher-order multipole moments of WFs beyond the quadrupole are often challenging to compute accurately. This approach allows the calculation to be broken down into independent Wannier-interpolated matrices. For example, for the QHC, instead of directly evaluating $\hat{P}_{b}(\hat{P}^{\alpha}_{a}\hat{P}^{\mu}_{a}\hat{P}^{\beta}_{b})^{(\text{W})}$, which would require a WFs octupole matrix, we compute $\hat{P}^{\alpha(\text{W})}_{a}$ and $(\hat{P}^{\mu}_{a}\hat{P}^{\beta}_{b})^{(\text{W})}$ separately, each involving at most dipole and quadrupole terms. 
    \par
    We use interQGT as an example to illustrate the derivation of such an expression from the general form:
    \begin{equation}
        \label{simplification of interQGT}
        \begin{aligned}
            Q^{\alpha\beta}_{ab} =& \mathrm{Tr}[\hat{P}_b \hat{P}^{\alpha}_{a}\hat{P}^{\beta}_{b}] \\
            =& \mathrm{Tr}[\hat{P}_b \hat{P}^{\alpha}_{a}\hat{P}_a\hat{P}^{\beta}_{b}] \\
            =& \sum_{jlmn} (\hat{P}^{(\text{H})}_{b})_{jl} \langle l|\hat{P}^{\alpha}_{a}|m\rangle
            (\hat{P}^{(\text{H})}_{a})_{mn} \langle n|\hat{P}^{\beta}_{b}|j\rangle \\
            =& \mathrm{Tr}[\hat{P}^{(\text{H})}_{b}\hat{P}^{\alpha(\text{W})}_{a}\hat{P}^{(\text{H})}_{a}\hat{P}^{\beta(\text{W})}_{b}].
        \end{aligned}
    \end{equation}
    Detailed procedures for evaluating other computable quantities in the Wannier representation are provided in Sec. II of the Supplementary Material. These methods can also be extended to intraband geometric quantities that related to transport properties.

    \subsection{Band Degeneracies}
    To conclude this section, we discuss in detail the complications associated with band degeneracies in the projector formalism. Although the projector method naturally allows for an analytical extension of geometric quantity calculations from the non-degenerate to the degenerate case, as demonstrated in Sec. \ref{Projector Formalism for Nonlinear Optics}, it remains challenging in numerical implementations. The main difficulty arises from the intrinsic difference between numerical and analytical treatments. In numerical calculations, the degenerate eigenstates of the Hamiltonian are obtained randomly at discrete $k$-points, with nondegenerate band indices assigned in ascending order of energy. This convention differs from that in analytical treatments, where eigenstates can be explicitly associated with specific band indices throughout the Brillouin zone. In the presence of degeneracies, this mismatch can lead to inconsistencies between the band labeling in numerical and analytical results.
    \par
    Such inconsistencies become particularly problematic when computing the derivatives of projectors using finite-difference schemes near degeneracy points. To illustrate this issue, consider two bands $a_1$ and $a_2$ that are degenerate at a momentum point $k' \in (k-dk, k+dk)$. Suppose the numerical and analytical labels coincide at $k - \delta k$ but are swapped at $k + \delta k$. In this case, the finite-difference approximation for the projector derivative, $\partial \hat{P}_{a_1}(k) \simeq (\hat{P}_{a_1}(k+dk) -\hat{P}_{a_1}(k-dk))/2dk$, involves projectors onto different analytic bands at $k + \delta k$ and $k - \delta k$. Since the corresponding eigenstates are nearly orthogonal, this mismatch can result in a spurious behavior of $\partial \hat{P}_{a_1}(k)$ in the vicinity of the degeneracy.
    \input{Figure/deg_consideration}
    \par
    To address these complications, additional care is required in the numerical treatment of degenerate bands. We introduce a degeneracy check by defining a small energy window $\delta\varepsilon$, within which $N_s$ bands are considered degenerate. Instead of computing quantities for individual bands, we directly evaluate the projector derivatives and geometric quantities associated with the degenerate subspace. This procedure avoids ambiguity related to band labelling. For example, consider the evaluation of a QHC. If two sets of bands, $a = \{a_1, \ldots, a_{N_s}\}$ and $b = \{b_1, \ldots, b_{M_s}\}$, form degenerate subspaces at a given $\boldsymbol{k}$-point, we define $\hat{P}_{a} = \sum_{a_s}^{N_s} \hat{P}_{a_s}$ and $\hat{P}_b= \sum_{b_s}^{M_s} \hat{P}_{b_s}$, and compute $C^{\mu;\alpha\beta}_{ab}(k)$ accordingly. To preserve a consistent total band count across the Brillouin zone, the resulting quantity is evenly distributed among the $N_s M_s$ band pairs as $C^{\mu;\alpha\beta}_{ab}(k)/(N_s M_s)$. This strategy effectively suppresses numerical instabilities near degeneracies and yields converged and smooth results in the Brillouin zone.
    \par

%% file: Table/Geometry_quantities_inter.tex
\begin{table*}[t]
    \centering
    \footnotesize
    \renewcommand{\arraystretch}{1.2}
    \begin{tabular}{c c c c}
        \toprule
        \toprule
        \textbf{Geometry} &  \textbf{General Expressions} & \textbf{WFs Expressions} & \textbf{Physical Phenomena} \\
        \midrule
        interQGT $Q^{\alpha\beta}_{ab}$ &
        $\text{Tr}\left[\hat{P}_b \hat{P}^{\alpha}_{a}\hat{P}^{\beta}_{b}\right]$
        & $ \text{Tr}\left[\hat{P}^{(\text{H})}_{b}\hat{P}^{\alpha{(\text{W})}}_{a}\hat{P}^{(\text{H})}_{a} \hat{P}^{\beta{(\text{W})}}_{b}\right]$ & \makecell*[c]{Injection Current \cite{BPVE_Sipe_2000, Injection_2nd}}\\
        [0.4cm]
        QHC $C^{\mu; \alpha\beta}_{ab}$ &
        $\text{Tr}\left[\hat{P}_b \hat{P}^{\alpha}_{a}\left(\hat{P}^{\mu\beta}_{b} + \hat{P}^{\mu}_{a}\hat{P}^{\beta}_{b}\right)\right]$
        &  $ \text{Tr}\left\{\hat{P}^{(\text{H})}_{b}\hat{P}^{\alpha{(\text{W})}}_{a}\hat{P}^{(\text{H})}_{a}\left[\hat{P}^{\mu\beta{(\text{W})}}_{b} + \left(\hat{P}^{\mu}_{a}\hat{P}^{\beta}_{b}\right)^{(\text{W})}\right]\right\}$ & \makecell*[c]{Shift Current \cite{Physical_illustration_of_shift_vector, CPL_Dai_2023}}\\
        [0.4cm]
        \makecell*[c]{TPP $T^{\mu\alpha\beta}_{abc}$} &
        \makecell*[c]{$\text{Tr}\left[\hat{P}_c \hat{P}^{\mu}_{a}\hat{P}^{\alpha}_{b}\hat{P}^{\beta}_{c}\right]$}
        &  \makecell*[c]{$\text{Tr}\left[\hat{P}^{(\text{H})}_c \hat{P}^{\mu{(\text{W})}}_{a}\hat{P}^{(\text{H})}_a\hat{P}^{\alpha{(\text{W})}}_{b}\hat{P}^{(\text{H})}_b\hat{P}^{\beta{(\text{W})}}_{c}\right]$} & \makecell*[c]{Third-Harmonic Generation\cite{THG_TPP}; \\Bicircular Light-induced \\ Photocurrent \cite{BCL}}\\
        [0.4cm]
        \bottomrule
        \bottomrule
    \end{tabular}
    \setlength{\abovecaptionskip}{0.4cm}
    \caption{General expressions for geometric quantities, their computable Wannier-representation  forms (WFs Expression) and corresponding physical phenomena. In the table, interQGT, TGT and TPP denote the interband quantum geometric tensor, quantum Hermitian connection and triple phase product, respectively.These interband quantities indicate that the subspace indices $a$ and $b$ (and $c$) are different. $(\hat{\mathcal{O}})^{(\text{W})}$ denotes the representation matrix of the operator $\hat{\mathcal{O}}$ in the Wannier basis.}
    \label{geometry quantities}
\end{table*}

%% file: Figure/deg_consideration.tex
\begin{figure}[ht]
    \centering
    \includegraphics[scale=0.6]{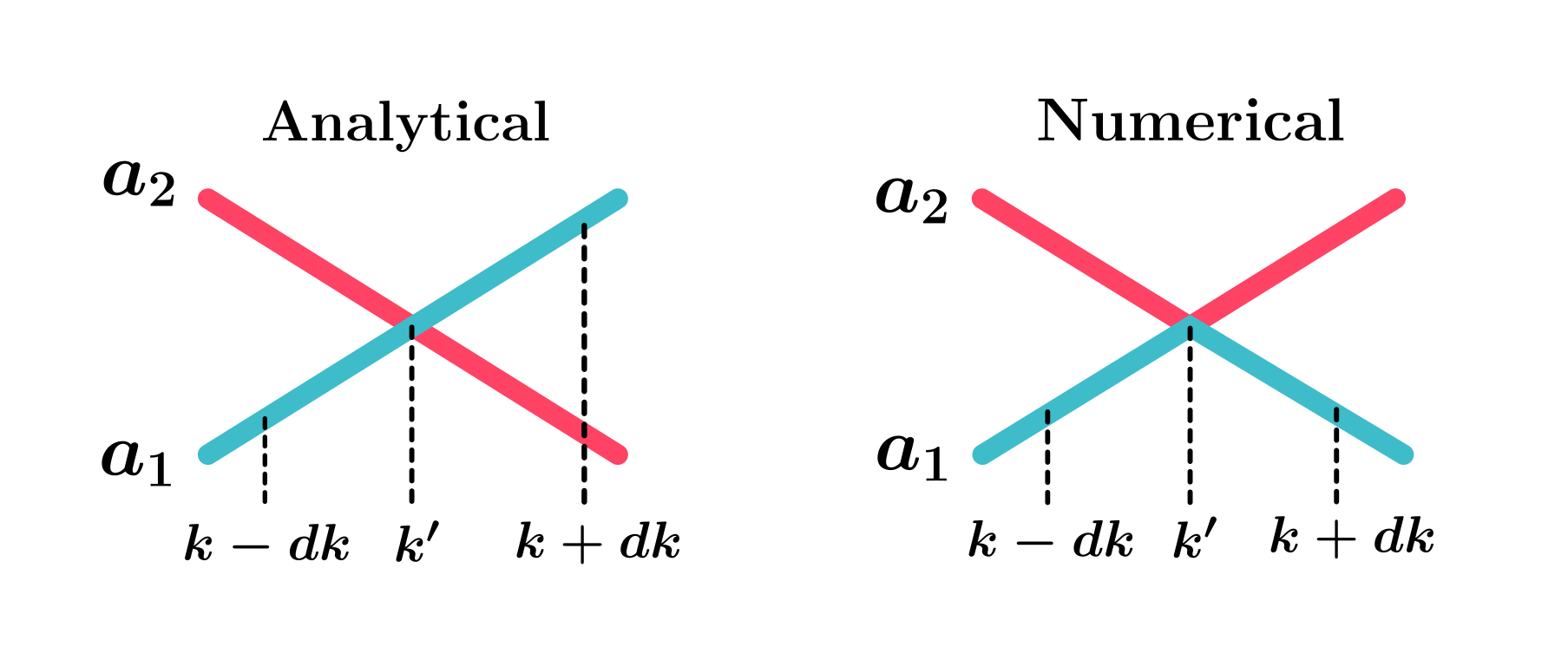}
    \caption{Analytical and numerical treatments near degeneracy points. “Analytical” and “Numerical” refer to the labeling of band indices in analytical and numerical calculations, respectively. The dashed lines connecting points on the bands indicate the bands actually involved when computing the finite-difference approximation of $\partial \hat{P}_{a_1}$.}
    \label{1st feyn diagram}
\end{figure}

%% file: sections/section04.tex
\section{Case Study: Shift Current of Monolayer GeS}
    \input{Figure/GeS_shift_current}
    \input{Figure/geometry_kspace}
    In this section, we assess the validity of our projector-formalism and Wannier interpolation based on first-principles calculations. Details of the computational procedure and parameter settings are provided in Sec. III of the Supplementary Material.
    \par
    As a representative example, we consider shift current in ferroelectric monolayer GeS, which exhibits a sizable shift current response under LPL \cite{GeS_shift, Sum_rule_Julen_2018}. The crystal and electronic band structures of monolayer GeS are shown in Fig. \ref{GeS shift current}(a) and (b), respectively. GeS belongs to the $C_{2v}$ point group and preserves a mirror symmetry $\mathcal{M}_x$ perpendicular to the $x$-axis. Symmetry considerations dictates that, for normally incident light propagating along the $z$ direction, only the second-order conductivity tensor components $\sigma^{yyy}_{\text{sh}}$, $\sigma^{yxx}_{\text{sh}}$, and $\sigma^{xxy}_{\text{sh}} = \sigma^{xyx}_{\text{sh}}$ are nonvanishing. In the following, we focus on the calculation of $\sigma^{yyy}_{\text{sh}}$, while results for the remaining components are provided in Sec. III of the Supplemental Material.
    \par
    We first illustrate the role of basis functions in the Wannier representation. Following the scheme outlined in Table~\ref{geometry quantities}, we compute the shift current by explicitly incorporating the full basis contributions. For comparison, we also evaluate the result using the projectors $\hat{P}^{\text{(H)}}$ in Hamiltonian gauge as components, in other word, neglecting the spatial spread of the basis functions. The corresponding results are presented in Fig. \ref{GeS shift current}(c). Evidently, whether the basis functions are neglected or not significantly affects the results, leading to discrepancies with purely first-principles calculations. This highlights the necessity of accounting for the finite spatial extent of the Wannier functions in these models.
    \par
    To benchmark the projector formalism, we directly compare it with two established methods, the sum rule method and the generalized Wilson loop approach. First, we briefly review how these two methods handle the shift current response. The core idea of the sum rule approach is to expand the generalized derivative $r^{\alpha}_{ab; \mu} = \partial_{\mu}r^{\alpha}_{ab} - i\left(r^{\beta}_{aa} - r^{\beta}_{bb} \right)r^{\alpha}_{ab}$, which is cumbersome to compute, as follows:
    \begin{equation}
        \begin{aligned}
            \label{sum rule}
            r^{\alpha}_{ab; \mu} =& \frac{i}{\varepsilon_{ab}}\left[\frac{v^{\alpha}_{ab}\Delta^{\beta}_{ab} +  v^{\beta}_{ab}\Delta^{\alpha}_{ab}}{\varepsilon_{ab}} - w^{\alpha\beta}_{ab} \right. \\
            &\left.+ \sum\limits^{N_{\text{wan}}}_{c\neq a, b}\left(\frac{v^{\alpha}_{ac}v^{\beta}_{cb}}{\varepsilon_{cb}} - \frac{v^{\beta}_{ac}v^{\alpha}_{cb}}{\varepsilon_{ac}}\right)\right].
        \end{aligned}
    \end{equation}
    Here, $v^{\alpha}_{ab} = \braket{a|\partial_{\alpha}\hat{H}|b}$ denotes the velocity matrix element, while $w^{\alpha\beta}_{ab} = \braket{a|\partial_{\alpha}\partial_{\beta}\hat{H}|b}$ represents the mass term \cite{mass_term}.  The sum rule approach introduces an auxiliary band index $c$, summing over all disentangled eigenstates $N_{\text{wan}}$, , which makes the calculation computationally expensive.  Additionally, this formulation suffers from numerical instabilities near band degeneracies, as $1/\varepsilon_{ac}$ diverges when bands $a$ and $c$ become nearly degenerate. To regularize this divergence, a small broadening parameter $\eta$ is typically introduced, replacing the singular factor with its principal-value form $\varepsilon_{ac}/(\varepsilon_{ac}^2 + \eta^2)$.
    We next discuss the generalized Wilson loop as a gauge-invariant method for evaluating geometric quantities\cite{Generalized_Wilson_Wang2022}. Within this formalism, the geometric quantity relevant to the shift current can be equivalently expressed through a gauge-invariant loop:
    \begin{equation}
        \begin{aligned}
            & -\partial_{q_{\mu}}W_{ba}(\boldsymbol{k}, \boldsymbol{q}, r^{\alpha}, \beta^{\beta}) \\
            &= -\partial_{q_{\mu}} \left[M^{\boldsymbol{k}; q_{\mu}}_{bb}r^{\alpha}_{ba}(\boldsymbol{k}+ q_{\mu})\left(M^{\boldsymbol{k}; q_{\mu}}_{aa}\right)^*r^{\alpha}_{ab}(\boldsymbol{k})\right],
        \end{aligned}
    \end{equation}
    where $M^{\boldsymbol{k}; q_{\mu}}_{bb} = \braket{b, \boldsymbol{k}|b, \boldsymbol{k}+ q_{\mu}}$is the overlap matrix. To connect with the projector formalism, we note that the interband Berry connection can be expressed as $r^{\alpha}_{ab} = iv^{\alpha}_{ab}/\varepsilon_{ba}$. Substituting this relation, one finds
    \begin{equation}
        \begin{aligned}
            \label{equivalence between wilson loop and projector}
            &-\partial_{q_{\mu}}W_{ba}(\boldsymbol{k}, \boldsymbol{q}, r^{\alpha}, \beta^{\beta}) \\
            &= \text{Tr}\left[\partial_{\mu}\left(\frac{\hat{P}_b\left(\partial_{\alpha}\hat{H}\right)\hat{P}_a}{\varepsilon_{ab}}\right)\frac{\hat{P}_a\left(\partial_{\beta}\hat{H}\right)\hat{P}_{b}}{\varepsilon_{ba}}\right]\\
            &= \text{Tr}\left[\partial_{\mu}\left(\hat{P}_b\hat{P}^{\alpha}_{a} \right)\hat{P}_a\hat{P}^{\beta}_b\right] \\
            &= \text{Tr}\left[\hat{P}_a \hat{P}^{\beta}_b\left(\hat{P}^{\mu\alpha}_{a} + \hat{P}^{\mu}_b \hat{P}^{\alpha}_{a}\right)\right] =C^{\mu, \alpha\beta}_{ba}.
        \end{aligned}
    \end{equation}
    This establishes the equivalence between the generalized Wilson loop and projector approaches. Nevertheless, their treatments of gauge invariance differ: the generalized Wilson loop transforms the entire geometric quantity into a gauge-invariant form before performing differentiation, whereas in the projector method, the projector itself is inherently gauge invariant. This difference in how gauge invariance is handled may lead to small numerical discrepancies between the two methods.
    \par
    The results obtained from these three methods are presented in Fig. \ref{GeS shift current}(d), showing an overall good agreement. In particular, the calculated conductivity exhibit excellent consistency in the low-frequency region, while slight discrepancies arise in the high-frequency regime. The discrepancies among the three methods originate from two main factors: (i) In practical calculations, applying the sum rule as in Eq. \ref{sum rule} requires the inclusion of an auxiliary band $c$. While the sum rule method is accurate when no bands are truncated, it tends to deviate slightly from the exact value in the high-energy region due to the truncation of bands. (ii) The three methods employ different procedures near degeneracy points. In the sum rule approach, a broadening parameter is introduced to avoid divergences caused by the inclusion of extra bands, but this also leads to deviations near degeneracies. The generalized Wilson loop approach and projector formalism, on the other hand, treat the bands degeneracies.
    \par
    At this stage, we have validated the projector method by reproducing the shift current conductivity. This method can be further extended to evaluate other NLO responses and multiband geometric quantities discussed in Sec.~\ref{Projector in Wannier Basis}. In Fig.~\ref{geometry kspace}, we present the results of three representative quantities associated with the conduction and valence bands near the Fermi surface of GeS: the interQGT, QHC, and TPP. These quantities can all be computed in the Wannier representation following the forms summarized in Table~\ref{geometry quantities}. The interQGT and QHC are relevant to the second-order injection and shift currents, respectively \cite{BPVE_Sipe_2000, Injection_2nd, Physical_illustration_of_shift_vector, CPL_Dai_2023}, while the TPP appears in third-order NLO responses and provides distinct multi-band contributions \cite{THG_TPP, BCL}.


%% file: Figure/GeS_shift_current.tex
\begin{figure*}[ht]
    \centering
    \includegraphics[scale=0.172]{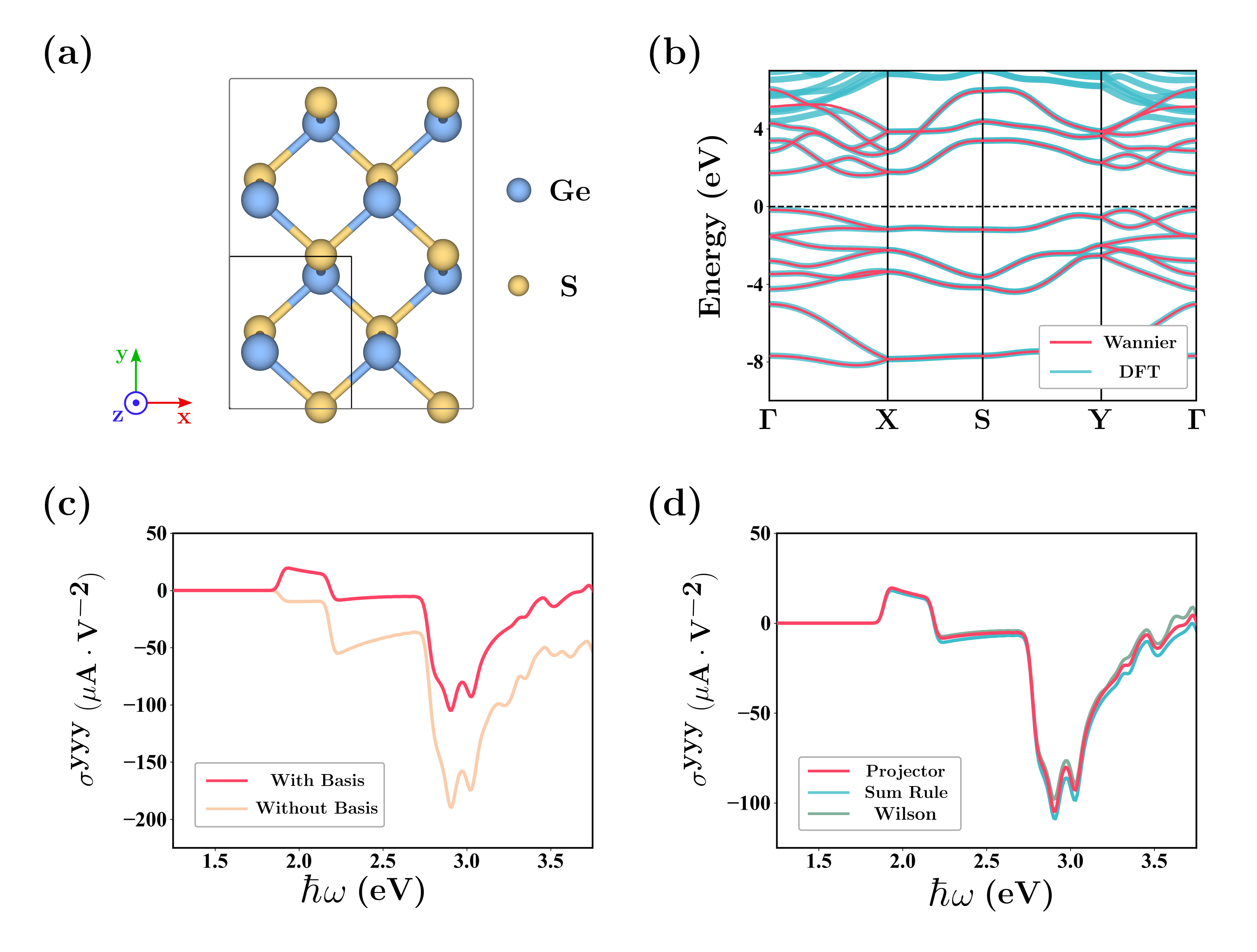}
    \caption{First-principles calculations of the shift current conductivity $\sigma^{yyy}$ in monolayer GeS. (a) Crystal structure of monolayer GeS, belonging to the $C_{2v}$ point group. (b) Band structures calculated by density functional theory (DFT) and Wannier interpolation. (c) Shift current conductivity calculated using the projector method, with (red) and without (yellow) the full basis broadening effect. (d) Shift current conductivity calculated by the projector method (red), the sum rule approach (blue), and the generalized Wilson loop method (green).}
    \label{GeS shift current}
\end{figure*}

%% file: Figure/geometry_kspace.tex
\begin{figure*}[ht]
    \centering
    \includegraphics[scale=0.172]{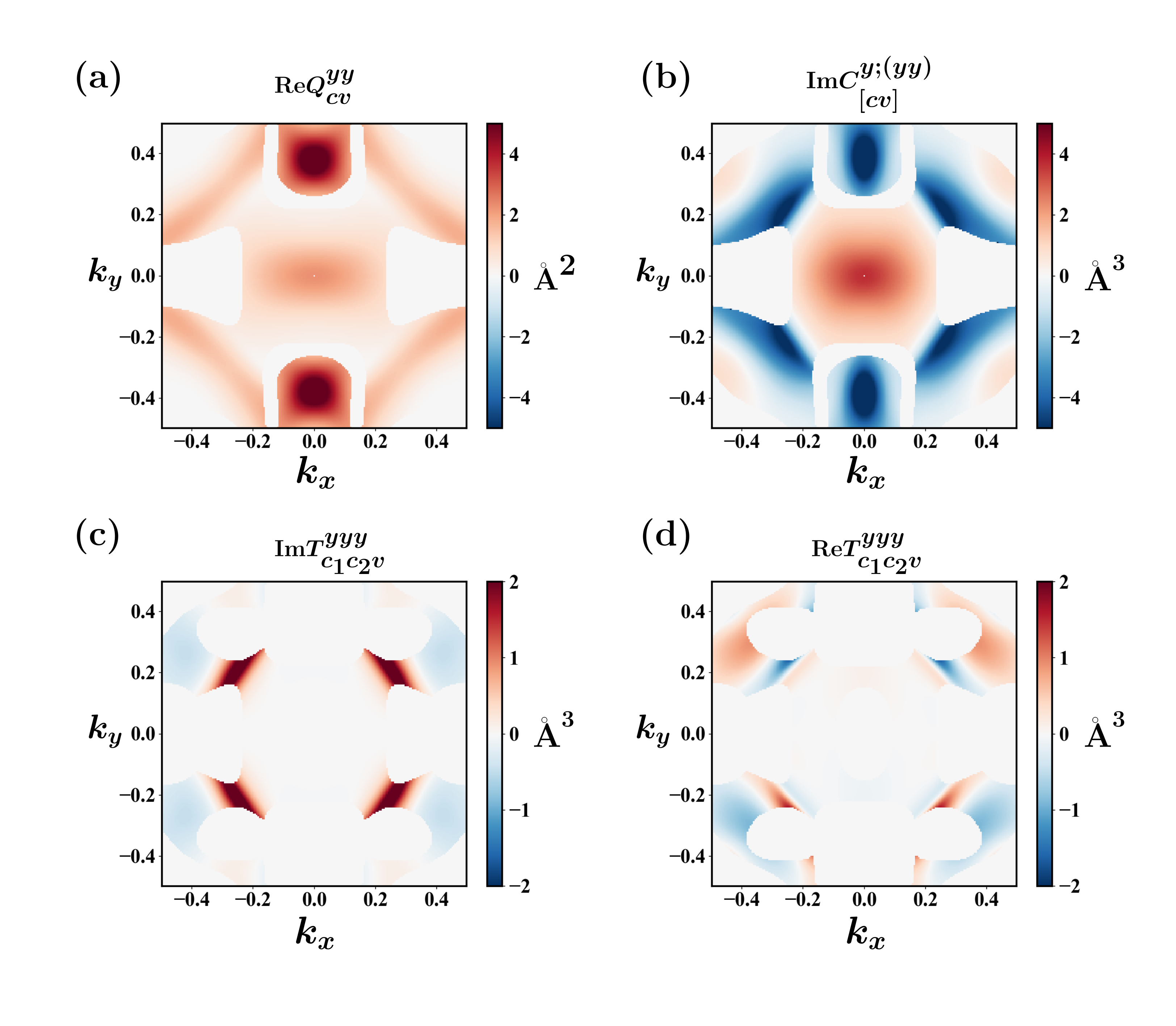}
    \caption{$\boldsymbol{k}$-resolved interband geometric quantities obtained from the projector method. (a) The $yy$ component of the quantum geometric tensor. (b) The imaginary part of the $yyy$ component of the third-order geometric tensor. (c, d) The imaginary and real parts of the $yyy$ component of the triple phase product. Panels (a,b) show two-band quantities involving only the conduction (c) and valence (v) bands nearest to the Fermi level, while panels (c,d) represent multi-band quantities including the two nearest conduction bands ($\text{c}_1$, $\text{c}_2$) and the nearest valence band (v).}
    \label{geometry kspace}
\end{figure*}

%% file: sections/section05.tex
\section{Discussion}
    In summary, we have developed a systematic framework based on the projector formalism to identify geometric quantities from Feynman diagrams and established a corresponding numerical scheme applicable to realistic materials, demonstrating both feasibility and distinct advantages. Using monolayer GeS as a representative example, we have shown that this approach alleviates certain intrinsic limitations of the conventional sum-rule method and reveals close connections with the generalized Wilson loop method. Additionally, we introduced a practical strategy for handling band degeneracies within the projector framework, enabling accurate evaluation of optical responses near degeneracy points. Owing to these favorable features, the projector formalism emerges as a robust and broadly applicable computational tool for NLO responses and quantum geometry. Future directions include extending this framework to other classes of NLO responses, such as the nonlinear anomalous Hall effect and photocurrents induced by bicircular light, as well as incorporating many-body effects and mixed geometric quantities, including electron–phonon coupling, excitonic effects, and polaronic contributions.

%% file: sections/acknowledgements.tex
\section*{Acknowledgements} \label{sec:acknowledgements}
    H.W. acknowledges the support from the NSFC under Grants Nos. 12474240, 12304049, and 12522411, as well as from the Zhejiang Provincial Natural Science Foundation under Grant No. LDT23F04014F01.

%% file: SM/SM01.tex
\section{Derivation of Resonant Second Order Response}
    In this section, we present in detail the computational procedure for deriving the resonant second-order response within the projector formalism based on Feynman diagrams. All Feynman diagrams contributing to the second-order response are summarized in Fig. \ref{SM 2nd feyn diagram}. We then proceed to illustrate how the resonant contribution can be systematically extracted and how the corresponding geometric quantities are expressed in the projector formalism.
    
    \input{Figure/SM_feynman_2nd}
    \subsection{Graph for Resonant Second Order Response}
    \label{Appendix:Graph for Resonant Second Order Response}
    Here, we describe how to extract the injection current and shift current from the Feynman diagrams of second-order NLO response. Both processes originate from resonant transitions, mathematically signaled by the presence of Dirac delta functions. In particular, the injection current scales with the relaxation time (or equivalently, with the inverse of the smearing factor). These features are encoded in the Green’s function products, so analyzing their structure allows us to identify the relevant Feynman diagrams. For second-order processes, all possible Green’s function products can be evaluated using contour integration techniques:
    \begin{equation}
        G^{(1)} = \int d \omega' G_a(\omega') = f_a ,
    \end{equation}
    \begin{equation}
        G^{(2)}(\Omega) = \int d \omega' G_a(\omega')G_b(\omega'+\Omega) = f_{ab} d^{\Omega}_{ab},
    \end{equation}
    \begin{equation}
        G^{(3)}(\Omega_1, \Omega_2) = \int d \omega' G_a(\omega')G_b(\omega'+\Omega_1)G_c(\omega'+\Omega_1 + \Omega_2) = f_{a} d^{\Omega_1}_{ab}d^{\Omega_{12}}_{ac} - f_{b} d^{\Omega_1}_{ab}d^{\Omega_2}_{bc} + f_{c} d^{\Omega_{12}}_{ac}d^{\Omega_2}_{bc}.
    \end{equation}
    Here we adopt the shorthand notation $\Omega_{12} = \Omega_1 + \Omega_2$. The superscript $(n)$ denotes the Green's function associated with the $n$-vertices diagram shown in Fig.~\ref{SM 2nd feyn diagram}.
    \par
    Consider monochromatic light induced BPVE. The input light frequencies $\omega_1$ and $\omega_2$ must equal $\pm \omega$. In this case, each type of Green's function product may have the following possible combinations of frequencies:
    \begin{equation}
        G^{(2)}: \Omega \in \{\omega, -\omega, 0\},
    \end{equation}
    \begin{equation}
        G^{(3)}: \{\Omega_1, \Omega_2\} \in \{\{\omega, -\omega\},\{-\omega, \omega\}\}.
    \end{equation}
    \par
    We first analyze the contribution to the injection current. Only $G^{(2)}(\Omega \neq 0)$ and $G^{(3)}$ can provide resonant contributions. The term proportional to the relaxation time does not appear in $G^{(2)}$; it arises only in $G^{(3)}$ when the subspace satisfies $\varepsilon_a=\varepsilon_c$, where $d^{\Omega_{12}}_{ac} = -i\gamma^{-1}$. The corresponding Green's function product is:
    \begin{equation}
        \begin{aligned}
            G^{(3)}_{a=c}(\Omega, -\Omega) =&  f_{a} d^{\Omega_1}_{ab}d^{0}_{ac} - f_{b} d^{\Omega_1}_{ab}d^{\Omega_2}_{bc} + f_{c} d^{\Omega_{0}}_{ac}d^{\Omega_2}_{bc} = -\frac{2\pi f_{ab}}{\gamma} \delta^{\Omega}_{ab} &(\Omega \in \{\omega, -\omega\}).
        \end{aligned}
    \end{equation}
    \par
    Next, we analyze the contribution to the shift current. Similar to the injection current, only $G^{(2)}(\Omega \neq 0)$ and $G^{(3)}$ can contribute to the shift current. Excluding the part associated with the injection current, all remaining Green's function products are independent of the relaxation time and therefore contribute to the shift current. The corresponding Green's function product are:
    \begin{equation}
        \begin{aligned}
            G^{(2)}(\Omega) =& f_{ab} d^{\Omega}_{ab} &(\Omega \in \{\omega, -\omega\}),
        \end{aligned}
    \end{equation}
    \begin{equation}
        \begin{aligned}
            G^{(3)}_{a\neq c}(\Omega, -\Omega) =& \frac{1}{\varepsilon_{ac}}\left(f_{ab} d^{\Omega}_{ab} - f_{bc} d^{\Omega}_{bc}\right) &\left(\Omega \in \{\omega, -\omega\}\right).
        \end{aligned}
    \end{equation}

    \subsection{Extracting Geometric Quantities}
    \label{Appendix:Extracting Geometric Quantities}
    In this subsection, we compute both currents within the projector formalism starting from the diagrams extracted above.
    \par
    \paragraph{Injection Current} According to the analysis above, the injection current is solely associated with contributions from three-vertex diagrams where the subspace satisfies $\varepsilon_a=\varepsilon_c$. The corresponding vertex product can be expressed as:
    \begin{equation}
        \begin{aligned}
            V^{\mu\alpha\beta, \text{inj}}_{ab} = \frac{e^3}{\hbar^3\omega^2}\text{Tr}\left[\hat{P}_a\left(\partial_{\mu}\hat{H}\right)\hat{P}_a\left(\partial_{\beta}\hat{H}\right)\hat{P}_b\left(\partial_{\alpha}\hat{H}\right)\hat{P}_a\right]
            = -\frac{e^3\varepsilon^2_{ab}}{\hbar^3\omega^2}\left(\partial_{\mu}\varepsilon_{a}\right) \text{Tr}\left[\hat{P}_a\hat{P}^{\beta}_b\hat{P}^{\alpha}_a\right]
            = -\frac{e^3\varepsilon^2_{ab}}{\hbar^3\omega^2}\left(\partial_{\mu}\varepsilon_{a}\right) Q^{\beta\alpha}_{ba},
        \end{aligned}
    \end{equation}
    where we define the quantum geometric tensor (interQGT) in projector formalism $Q^{\beta\alpha}_{ba} \equiv \text{Tr}\left[\hat{P}_a\hat{P}^{\beta}_b\hat{P}^{\alpha}_a\right]$. By supplementing the vertex product with the Green's function product and the coefficient product, we obtain the conductivity of the injection current:
    \begin{equation}
        \begin{aligned}
            \sigma^{\mu\alpha\beta}_{\text{inj}} =& \frac{e^3}{\hbar^3\omega^2}\int[d\boldsymbol{k}]\sum\limits_{a, b}\varepsilon^2_{ab}\left(\partial_{\mu}\varepsilon_{a}\right)Q^{\beta\alpha}_{ba}\frac{2\pi f_{ab}}{\gamma}\delta^{\omega}_{ab} +  \left[(\alpha, \beta, \omega)\leftrightarrow(\beta, \alpha, -\omega)\right] \\
            =& -\frac{2\pi e^3}{\gamma}\int[d\boldsymbol{k}]\sum\limits_{a, b} f_{ab}\Delta^{\mu}_{ab} Q^{\alpha\beta}_{ab}\delta^{\omega}_{ab},
        \end{aligned}
    \end{equation}
    where we defined $\Delta^{\mu}_{ab} = \partial_{\mu} \varepsilon_{ab}$ as the group velocity difference between subspace $a$ and $b$. In time-reversal symmetric systems, the injection current is purely imaginary, and thus can be induced only by circularly polarized light (CPL).
    \paragraph{Shift Current} In the case of the shift current, the relevant Feynman diagrams involve both first- and second-order derivatives of the Hamiltonian. The vertex products associated with the two-vertices and three-vertices diagrams take the form
    \begin{equation}
        V^{\mu\alpha\beta, \text{sh(2)}}_{ab, a\neq b} = \frac{e^3}{\hbar^3\omega^2}\text{Tr}\left[\hat{P}_b\left(\partial_{\alpha}\hat{H}\right)\hat{P}_{a}\left(\partial_{\mu}\partial_{\beta}\hat{H}\right)\hat{P}_b\right],
    \end{equation}
    \begin{equation}
        V^{\mu\alpha\beta, \text{sh(3)}}_{abc, c\neq a} = \frac{e^3}{\hbar^3\omega^2}\text{Tr}\left[\hat{P}_b\left(\partial_{\alpha}\hat{H}\right)\hat{P}_{a}\left(\partial_{\mu}\hat{H}\right)\hat{P}_{c}\left(\partial_{\beta}\hat{H}\right)\hat{P}_b\right].
    \end{equation}
    By applying the expansion rules, these vertex products can be expressed as
    \begin{equation}
        \begin{aligned}
            \label{expanded V2 shift}
            V^{\mu\alpha\beta, \text{sh(2)}}_{ab, a\neq b} =& \frac{e^3}{\hbar^3\omega^2}\left\{-\varepsilon^2_{ab} \text{Tr}\left[\hat{P}_b\hat{P}^{\alpha}_a\left(\hat{P}^{\mu\beta}_b + \hat{P}^{\mu}_a\hat{P}^{\beta}_b\right)\right]+ \varepsilon_{ab}\partial_{\mu}\varepsilon_{ba}\text{Tr}\left[\hat{P}_b \hat{P}^{\alpha}_a\hat{P}^{\beta}_b\right]+ \varepsilon_{ab}\partial_{\beta}\varepsilon_{ba}\text{Tr}\left[\hat{P}_b\hat{P}^{\alpha}_a \hat{P}^{\mu}_b\right]\right. \\
            &\left. + \sum\limits_{c \neq a \neq b}\varepsilon_{ab}\varepsilon_{ac}\text{Tr}\left[\hat{P}_b\hat{P}^{\alpha}_a \hat{P}^{\beta}_c\hat{P}^{\mu}_b\right] - \sum\limits_{c \neq a \neq b}\varepsilon_{ab}\varepsilon_{cb}\text{Tr}\left[\hat{P}_b\hat{P}^{\alpha}_a\hat{P}^{\mu}_c\hat{P}^{\beta}_b\right]\right\},
        \end{aligned}
    \end{equation}
    \begin{equation}
        \begin{aligned}
            \label{expanded V3 shift}
            V^{\mu\alpha\beta, \text{sh(3)}}_{abc, c\neq a} =& \frac{e^3}{\hbar^3\omega^2}\left\{-\delta_{ab}\varepsilon^2_{ac}\partial_{\alpha}\varepsilon_{a}\text{Tr}\left[\hat{P}_a \hat{P}^{\mu}_c\hat{P}^{\beta}_a\right] - \delta_{bc}\varepsilon^2_{ab}\partial_{\beta}\varepsilon_{b}\text{Tr}\left[\hat{P}_b \hat{P}^{\alpha}_a\hat{P}^{\mu}_b\right] \right.\\
            &\left.+ (1-\delta_{ac})(1-\delta_{bc})\varepsilon_{ab}\varepsilon_{ca}\varepsilon_{bc}\text{Tr}\left[\hat{P}_b\hat{P}^{\alpha}_a\hat{P}^{\mu}_c\hat{P}^{\beta}_b\right]\right\}.
        \end{aligned}
    \end{equation}
    
    Following the Feynman rules, the full conductivity expression is obtained by multiplying the vertex product with the corresponding Green’s function product. After summing the contributions from both diagrams, it can be shown that the terms from the second and third lines of Eq.~\ref{expanded V2 shift} (the two-vertex diagram) exactly cancel those from Eq.~\ref{expanded V3 shift} (the three-vertex diagram). The corresponding conductivities are denoted as $\sigma^{\mu\alpha\beta \text{(2')}}_{\text{sh}}$ and $\sigma^{\mu\alpha\beta \text{(3)}}_{\text{sh}}$, respectively, with explicit expressions given by:
    \begin{equation}
        \begin{aligned}
            \sigma^{\mu\alpha\beta \text{(2')}}_{\text{sh}} =& \frac{e^3}{\hbar^3\omega^2} \int[d\boldsymbol{k}]\sum\limits_{ab} f_{ab}d^{\omega}_{ab} \left\{\varepsilon_{ab}\partial_{\beta}\varepsilon_{ba}\text{Tr}\left[\hat{P}_b \hat{P}^{\alpha}_a \hat{P}^{\mu}_b\right] + \sum\limits_{c \neq a \neq b}\varepsilon_{ab}\varepsilon_{ac}\text{Tr}\left[\hat{P}_b\hat{P}^{\alpha}_a\hat{P}^{\beta}_c \hat{P}^{\mu}_b\right] \right.\\
            &\left.- \sum\limits_{c \neq a \neq b}\varepsilon_{ab}\varepsilon_{cb}\text{Tr}\left[\hat{P}_b\hat{P}^{\alpha}_a \hat{P}^{\mu}_c \hat{P}^{\beta}_b\right]\right\} + \left[(\alpha, \beta, \omega) \leftrightarrow (\beta, \alpha, -\omega)\right],
        \end{aligned}
    \end{equation}
    \begin{equation}
        \begin{aligned}
            \sigma^{\mu\alpha\beta \text{(3)}}_{\text{sh}} =& \frac{e^3}{\hbar^3\omega^2} \int[d\boldsymbol{k}]\sum\limits_{abc, a\neq c} \left(\frac{f_{ab}}{\varepsilon_{ac}}d^{\omega}_{ab} - \frac{f_{bc}}{\varepsilon_{ac}}d^{-\omega}_{bc}\right) \left\{-\delta_{ab}\varepsilon^2_{ac}\partial_{\alpha}\varepsilon_{a}\text{Tr}\left[\hat{P}_a \hat{P}^{\mu}_c \hat{P}^{\beta}_a\right] \right.\\
            &\left.- \delta_{bc}\varepsilon^2_{ab}\partial_{\beta}\varepsilon_{b}\text{Tr}\left[\hat{P}_b \hat{P}^{\alpha}_a \hat{P}^{\mu}_b\right] + (1-\delta_{ac})(1-\delta_{bc})\varepsilon_{ab}\varepsilon_{ca}\varepsilon_{bc}\text{Tr}\left[\hat{P}_b\hat{P}^{\alpha}_a\hat{P}^{\mu}_c\hat{P}^{\beta}_b\right]\right\}  \\
            &+ \left[(\alpha, \beta, \omega) \leftrightarrow (\beta, \alpha, -\omega)\right].
        \end{aligned}
    \end{equation}
    After rearranging the indices and exploiting the intrinsic properties of the projectors, one finds that the two contributions are exact negatives of each other:
    \begin{equation}
        \begin{aligned}
            \sigma^{\mu\alpha\beta \text{(3)}}_{\text{sh}} =& \frac{e^3}{\hbar^3\omega^2} \int[d\boldsymbol{k}]\sum\limits_{a\neq b \neq c}\left(f_{bc}d^{-\omega}_{bc} - f_{ab}d^{\omega}_{ab} \right)\varepsilon_{ab}\varepsilon_{bc}\text{Tr}\left[\hat{P}_b \hat{P}^{\alpha}_a \hat{P}^{\mu}_c\hat{P}^{\beta}_b\right] \\
            & +\sum\limits_{ac, a\neq c}f_{ac}d^{-\omega}_{ac}\varepsilon_{ac}\partial_{\alpha}\varepsilon_{a} \text{Tr}\left[\hat{P}_a \hat{P}^{\mu}_c \hat{P}^{\beta}_a\right] - \sum\limits_{ab, a\neq b}f_{ab}d^{\omega}_{ab}\varepsilon_{ab}\partial\varepsilon_{b}\text{Tr}\left[\hat{P}_b \hat{P}^{\alpha}_a\hat{P}^{\mu}_b\right]\\
            &+ \left[(\alpha, \beta, \omega) \leftrightarrow (\beta, \alpha, -\omega)\right] \\
            =& -\frac{e^3}{\hbar^3\omega^2} \int[d\boldsymbol{k}]\sum\limits_{ab} f_{ab}d^{\omega}_{ab} \left\{\varepsilon_{ab}\partial_{\beta}\varepsilon_{ba}\text{Tr}\left[\hat{P}_b \hat{P}^{\alpha}_a \hat{P}^{\mu}_b\right] \right.+ \sum\limits_{c \neq a \neq b}\varepsilon_{ab}\varepsilon_{ac}\text{Tr}\left[\hat{P}_b \hat{P}^{\alpha}_a \hat{P}^{\beta}_c \hat{P}^{\mu}_b\right]\\
            &\left. - \sum\limits_{c \neq a \neq b}\varepsilon_{ab}\varepsilon_{cb}\text{Tr}\left[\hat{P}_b \hat{P}^{\alpha}_a\hat{P}^{\mu}_c\hat{P}^{\beta}_b\right]\right\} + \left[(\alpha, \beta, \omega) \leftrightarrow (\beta, \alpha, -\omega)\right]\\
            =& - \sigma^{\mu\alpha\beta \text{(2')}}_{\text{sh}}.
        \end{aligned}
    \end{equation}
    
    Consequently, only the remaining terms in Eq.~\ref{expanded V2 shift} contribute to the conductivity. We denote this part as $\sigma^{\mu\alpha\beta}_{\text{sh}}$. We then extract the resonant contribution, which dominates the conductivity. In practical calculations, this is achieved by replacing all $d$-functions with their imaginary parts, e.g. $d^{\omega}_{ab} \rightarrow -i\pi\delta^{\omega}_{ab}$.
    \begin{equation}
        \begin{aligned}
            \label{projector 2nd shift}
            \sigma^{\mu\alpha\beta}_{\text{sh}} =& i\pi \frac{e^3}{\hbar^3\omega^2} \int[d\boldsymbol{k}]\sum\limits_{ab} f_{ab}\delta^{\omega}_{ab}\varepsilon^2_{ab} \text{Tr}\left[\hat{P}_b\hat{P}^{\alpha}_a\left(\hat{P}^{\mu\beta}_b + \hat{P}^{\mu}_a\hat{P}^{\beta}_b\right)\right] \\
            &- f_{ab}\delta^{\omega}_{ab}\varepsilon_{ab}\partial_{\mu}\varepsilon_{ba}\text{Tr}\left[\hat{P}_b \hat{P}^{\alpha}_a \hat{P}^{\beta}_b\right] + \left[(\alpha, \beta, \omega) \leftrightarrow (\beta, \alpha, -\omega)\right] \\
            =& \frac{i\pi e^3}{\hbar^2} \int[d\boldsymbol{k}]\sum\limits_{ab} f_{ab}\delta^{\omega}_{ab} \left(C^{\mu;\alpha\beta}_{ab} - C^{\mu;\alpha\beta}_{ba}\right) -\frac{f_{ab}}{\varepsilon_{ab}}\delta^{\omega}_{ab}\left(\partial_{\mu}\varepsilon_{ba} - \partial_{\mu}\varepsilon_{ba}\right)Q^{\alpha\beta}_{ab} \\
            =&\frac{i\pi e^3}{\hbar^2} \int[d\boldsymbol{k}]\sum\limits_{ab} f_{ab}\delta^{\omega}_{ab} \left(C^{\mu;\alpha\beta}_{ab} - C^{\mu;\alpha\beta}_{ba}\right).
        \end{aligned}
    \end{equation}
    The quantity $C^{\mu; \alpha\beta}_{ab} = \text{Tr}\left[\hat{P}_b \hat{P}^{\alpha}_a\left(\hat{P}^{\mu\beta}_b + \hat{P}^{\mu}_a \hat{P}^{\beta}_b\right)\right]$ is a quantum hermitian connection (QHC). As a result, only the terms involving the QHC contribute to the second-order shift current.
    \par
    Further considering the shift current induced by linearly polarized light (LPL), this photocurrent constitutes the sole contribution to the resonant second order response in time-reversal symmetric systems. The corresponding conductivity is given by the real part of Eq.~\ref{projector 2nd shift}:
    \begin{equation}
        \begin{aligned}
            \sigma^{\mu\alpha\beta}_{\text{sh, LPL}} =& \text{Re}\left[\sigma^{\mu\alpha\beta}_{\text{sh}}\right]
            = \frac{\pi e^3}{\hbar^2} \int[d\boldsymbol{k}]\sum\limits_{ab} f_{ab}\delta^{\omega}_{ab} \text{Im}C^{\mu;(\alpha\beta)}_{[ab]}.
        \end{aligned}
    \end{equation}
    Here we define $(\alpha\beta)$ and $[ab]$ as the symmetric and antisymmetric brackets, respectively, i.e. $C^{\mu;(\alpha\beta)}_{[ab]} = (C^{\mu;\alpha\beta}_{ab}+C^{\mu;\beta\alpha}_{ab} - C^{\mu;\alpha\beta}_{ba}- C^{\mu;\beta\alpha}_{ba})/4$.

%% file: Figure/SM_feynman_2nd.tex
\begin{figure}[ht]
    \centering
    \includegraphics[scale=0.6]{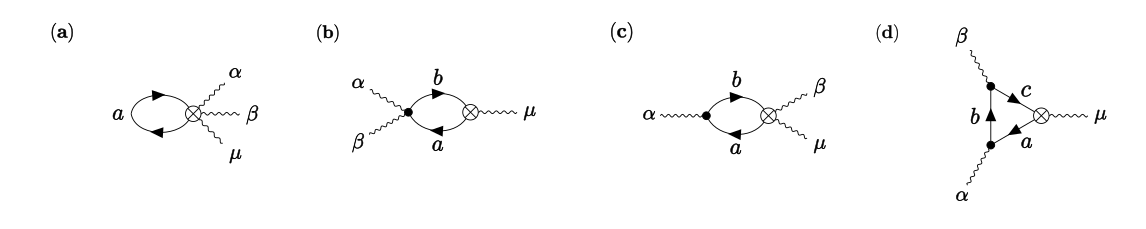}
    \caption{Feynman diagram corresponding to the second order response. (a), (b, c) and (d) corresponds to one-vertex, two-vertices and three-vertices diagram, respectively. The indices ${\alpha, \beta}$ and ${\mu}$ denote the Cartesian components of the three incident light fields and the output current, respectively, while ${a, b, c}$ label the band indices of each Green’s function.}
    \label{SM 2nd feyn diagram}
\end{figure}

%% file: SM/SM02.tex
\section{Derivation of Projector formalism in Wannier Basis}
    \subsection{Projector Derivative}
    We begin by supplementing the derivation of the first- and second-order derivatives of the projector in the Wannier basis. The first order derivative, $\partial_{\alpha}P^{(\text{W})}_{n} \equiv P^{\alpha (\text{W})}_{n}$, reads
    \begin{equation}
        \begin{aligned}
            \label{wannier 1st diff}
            \left(P^{\alpha (\text{W})}_{n}\right)_{ij} =& \bra{i}\partial_{\alpha}\left(\ket{u_n}\bra{u_n}\right)\ket{i} \\
            =& \sum\limits_{l, m}\partial_{\alpha}\left(U^{\dagger}_{ln}U_{nm}\right)\braket{i|l}\braket{m|j} + \left(U^{\dagger}_{ln}U_{nm}\right)\braket{i|l}\braket{m|\partial_{\alpha}j} + \partial_{\alpha}\left(U_{ln}U^{\dagger}_{nm}\right)\braket{i|l}\braket{\partial_{\alpha}m|j} \\
            =& \left(P^{\alpha (\text{H})}_{n}\right)_{ij} + \sum\limits_{l}\left(P^{(\text{H})}_{n}\right)_{lj}\braket{i|\partial_{\alpha}l} - \sum\limits_{m} \left(P^{(\text{H})}_{n}\right)_{im}\braket{m|\partial_{\alpha}j} \\
            =& \left(P^{\alpha (\text{H})}_{n}\right)_{ij} - \sum\limits_{l}i\left(P^{(\text{H})}_{n}\right)_{lj}A^{\alpha }_{il} + \sum\limits_{m} i\left(P^{(\text{H})}_{n}\right)_{im}A^{\alpha}_{mj} \\
            \neq & \left(P^{\alpha (\text{H})}_n\right)_{ij},
        \end{aligned}
    \end{equation}
    where $A^{\alpha}_{mj}(\boldsymbol{k})$denotes the Fourier transform of the dipole matrix, $A^{\alpha}_{mj}(\boldsymbol{k}) = \braket{m|i\partial_{\alpha}j} = \sum_{\boldsymbol{R}}e^{i\boldsymbol{k}\boldsymbol{R}}\braket{\boldsymbol{0}m|\hat{r_{\alpha}}|\boldsymbol{R}j}$.
    \par
    Similarly, the second order derivative $\partial_{\alpha} \partial_{\beta} P^{(\text{W})}_n \equiv P^{\alpha\beta(\text{W})}_n $ is given by
    \begin{equation}
        \begin{aligned}
            \label{wannier 2nd diff}
            \left(P^{\alpha\beta(\text{W})}_n\right)_{ij} =& \bra{i} \partial_{\alpha}\partial_{\beta}\left(\ket{n}\bra{n}\right)\ket{j} \\
            =& \sum\limits_{l, m} \bra{i} \left(\partial_{\alpha}\partial_{\beta} U^{\dagger}_{ln}U_{nm}\ket{l}\bra{m}\right) \ket{j} \\
            =& \sum\limits_{l,m} \bra{i}\left[\left(P^{\alpha\beta(\text{H})}_n\right)_{lm}\ket{l}\bra{m} + \left(P^{\beta(\text{H})}_n\right)_{lm}\ket{\partial_{\alpha}l}\bra{m} + \left(P^{\beta(\text{H})}_n\right)_{lm}\ket{l}\bra{\partial_{\alpha}m} \right. \\
            &+ \left(P^{\alpha(\text{H})}_n\right)_{lm}\ket{\partial_{\beta}l}\bra{m} + \left(P^{(\text{H})}_n\right)_{lm}\ket{\partial_{\alpha}\partial_{\beta}l}\bra{m} + \left(P^{(\text{H})}_n\right)_{lm}\ket{\partial_{\beta}l}\bra{\partial_{\alpha}m}  \\
            &\left. + \left(P^{\alpha(\text{H})}_n\right)_{lm}\ket{l}\bra{\partial_{\beta}m} + \left(P^{(\text{H})}_n\right)_{lm}\ket{\partial_{\alpha}l}\bra{\partial_{\beta}m} + \left(P^{(\text{H})}_n\right)_{lm}\ket{l}\bra{\partial_{\alpha}\partial_{\beta}m}\right]\ket{j} \\
            =& \left(P^{\alpha\beta(\text{H})}_n\right)_{ij} + \sum\limits_{l} \left[-i\left(P^{\beta(\text{H})}_n\right)_{lj}A^{\alpha}_{il}-i\left(P^{\alpha(\text{H})}_n\right)_{lj}A^{\beta}_{il} - \left(P^{(\text{H})}_n\right)_{lj}D^{\dagger, \alpha\beta}_{il} \right] \\
            & + \sum\limits_{m} \left[i\left(P^{\beta(\text{H})}_n\right)_{im}A^{\alpha}_{mj}+i\left(P^{\alpha(\text{H})}_n\right)_{im}A^{\beta}_{mj} - \left(P^{(\text{H})}_n\right)_{im}D^{\alpha\beta}_{mj} \right] \\
            & + \sum\limits_{l,m} \left[\left(P^{(\text{H})}_n\right)_{lm}A^{\alpha}_{il}A^{\beta}_{mj} + \left(P^{(\text{H})}_n\right)_{lm}A^{\beta}_{il}A^{\alpha}_{mj} \right] \\
            \neq & \left(P^{\alpha\beta(\text{H})}_n\right)_{ij},
        \end{aligned}
    \end{equation}
    where $D^{\alpha\beta}_{ij}$ is the quadrupole matrix in Wannier basis, i.e. $D^{\alpha\beta}_{ij} = -\braket{\partial_{\alpha}\partial_{\beta}i|j} = \sum_{\boldsymbol{R}}e^{i\boldsymbol{k}\boldsymbol{R}}\braket{\boldsymbol{0}i|\hat{r}_{\alpha}\hat{r}_{\beta}|\boldsymbol{R}j}$.
    \par
    Besides the above two formulas, the calculation of the shift current within the projector formalism also requires an additional expression for the component $\left(\hat{P}^{\alpha}_{m}\hat{P}^{\beta}_{n}\right)^{(\text{W})}$.
        \begin{equation}
            \begin{aligned}
                &\braket{i|\left(\hat{P}^{\alpha}_{m}\hat{P}^{\beta}_{n}\right)^{(\text{W})}|j} \\
                =&\sum\limits_{l, m, p, q} \braket{i|\partial_{\alpha}\left(\ket{l}\left(\hat{P}^{(\text{H})}_{a}\right)_{lm}\bra{m}\right)\partial_{\beta}\left(\ket{p}\left(\hat{P}^{(\text{H})}_{b}\right)_{pq}\bra{q}\right)|j} \\
                =&\sum\limits_{l}\left(\hat{P}^{\alpha(\text{H})}_{a}\right)_{il}\left(\hat{P}^{\beta(\text{H})}_{b}\right)_{lj} + i \sum\limits_{lm}\left[\left(\hat{P}^{\alpha(\text{H})}_{a}\right)_{il}A^{\beta}_{lm}\left(\hat{P}^{(\text{H})}_{b}\right)_{mj}   - \left(\hat{P}^{\alpha(\text{H})}_{a}\right)_{il}\left(\hat{P}^{(\text{H})}_{b}\right)_{lm} A^{\beta}_{mj} \right.\\
                &\left.- A^{\alpha}_{il}\left(\hat{P}^{(\text{H})}_a\right)_{lm}\left(\hat{P}^{\beta(\text{H})}_{b}\right)_{mj}   + \left(\hat{P}^{(\text{H})}_a\right)_{il}A^{\alpha}_{lm}\left(\hat{P}^{\beta(\text{H})}_{b}\right)_{mj} \right]\\
                &+\sum\limits_{lmn}\left[A^{\alpha}_{il}\left(\hat{P}^{(\text{H})}_a\right)_{lm}\left(\hat{P}^{(\text{H})}_b\right)_{mn}A^{\beta}_{nj} - A^{\alpha}_{il}\left(\hat{P}^{(\text{H})}_a\right)_{lm}A^{\beta}_{mn}\left(\hat{P}^{(\text{H})}_b\right)_{nj} \right.\\
                &\left.- \left(\hat{P}^{(\text{H})}_a\right)_{il}A^{\alpha}_{lm}\left(\hat{P}^{(\text{H})}_b\right)_{mn}A^{\beta}_{nj} + \left(\hat{P}^{(\text{H})}_a\right)_{il}D^{\alpha\beta'}_{lm}\left(\hat{P}^{(\text{H})}_b\right)_{mj}\right].
            \end{aligned}
        \end{equation}
    Here we define $D^{\alpha\beta'}_{lm} = \braket{\partial_{\alpha} l|\partial_{\beta}m} = \sum\limits_{\boldsymbol{R}}e^{i\boldsymbol{k R}}\braket{\boldsymbol{0}l|\hat{r}_{\alpha}(\hat{r}_{\beta} - R_{\beta})|\boldsymbol{R}m}$.
    \par
        
    \subsection{Quantum Geometry in Wannier representation}
    Based on the projector derivatives obtained above, we transform several important interband geometric quantities into computable Wannier-representation forms. While the main text has already illustrated the transformation procedure for the interband quantum geometric tensor (interQGT), here we present the corresponding formulations for the QHC $C^{\mu; \alpha\beta}_{ab}$ and triple phase product (TPP) $T^{\mu\alpha\beta}_{abc}$:
    \begin{equation}
        \begin{aligned}
            C^{\mu; \alpha\beta}_{ab} =& \text{Tr}\left[\hat{P}_{b}\hat{P}^{\alpha}_{a}\left(\hat{P}^{\mu\beta}_{b} + \hat{P}^{\mu}_{a}\hat{P}^{\beta}_{b}\right)\right] = \text{Tr}\left[\hat{P}_{b}\hat{P}^{\alpha}_{a} \hat{P}_{a}\left(\hat{P}^{\mu\beta}_{b} + \hat{P}^{\mu}_{a}\hat{P}^{\beta}_{b}\right)\right] \\
            =& \sum\limits_{jlmn} \left(\hat{P}^{(\text{H})}_{a}\right)_{jl}\braket{l|\hat{P}^{\alpha}_{a}|m}\left(\hat{P}^{(\text{H})}_{b}\right)_{mn}\left(\braket{n|\hat{P}^{\mu\beta}_{b}|j}+\braket{n|\hat{P}^{\mu}_{b}\hat{P}^{\alpha}_{b}|j}\right) \\
            =& \text{Tr}\left\{\hat{P}^{(\text{H})}_{b}\hat{P}^{\alpha(\text{W})}_{a}\hat{P}^{(\text{H})}_{a}\left[\hat{P}^{\mu\beta(\text{W})}_{b} + \left(\hat{P}^{\mu}_{a}\hat{P}^{\beta}_{b}\right)^{(\text{W})}\right]\right\},
        \end{aligned}
    \end{equation}
    \begin{equation}
        \begin{aligned}
            T^{\mu\alpha\beta}_{abc} = & \text{Tr}\left[\hat{P}_c \hat{P}^{\mu}_{a}\hat{P}^{\alpha}_{b}\hat{P}^{\beta}_{c}\right]
            = \text{Tr}\left[\hat{P}_c \hat{P}^{\mu}_{a}\hat{P}_{a}\hat{P}^{\alpha}_{b}\hat{P}_{b}\hat{P}^{\beta}_{c}\hat{P}_{c}\right] \\
            =& \sum\limits_{jlmnop} \left(\hat{P}^{(\text{H})}_{c}\right)_{jl}\braket{l|\hat{P}^{\mu}_{a}|m}\left(\hat{P}^{(\text{H})}_{a}\right)_{mn}\braket{n|\hat{P}^{\alpha}_{b}|o}\left(\hat{P}^{(\text{H})}_{b}\right)_{op}\braket{p|\hat{P}^{\beta}_{c}|j} \\
            =& \text{Tr}\left[\hat{P}^{(\text{H})}_c \hat{P}^{\mu(\text{W})}_{a}\hat{P}^{(\text{H})}_a\hat{P}^{\alpha(\text{W})}_{b}\hat{P}^{(\text{H})}_b\hat{P}^{\beta(\text{W})}_{c}\right].
        \end{aligned}
    \end{equation}

%% file: SM/SM03.tex
\section{First-Principles Calculations}
\label{appendix: numerical calculation details}
    In this section, we provide a detailed description of the first-principles simulation setup for GeS, and specify the parameter selection procedures for calculating the shift current within the projector formalism.
    \par
    The Wannier tight-binding (TB) model for GeS was constructed following the standard DFT+Wannier90 workflow. First-principles calculations were performed using the QUANTUM ESPRESSO package within the self-consistent field (SCF) framework. A monolayer GeS supercell was adopted, and a vacuum layer was introduced along the non-periodic $z$-direction to eliminate spurious interlayer interactions, with the out-of-plane lattice constant set to 25.4~\text{\AA}. The plane-wave basis was truncated at a kinetic energy cutoff of 60 Ry, and Brillouin zone sampling was performed using a $12 \times 10 \times 1$ Monkhorst-Pack $k$-point grid. A total of 72 bands near the Fermi level were computed. Norm-conserving pseudopotentials were employed to describe the core-valence interaction, and exchange-correlation effects were treated within the Perdew–Burke–Ernzerhof (PBE) generalized gradient approximation. The WANNIER90 package was then used to construct maximally localized Wannier functions (MLWFs). Based on the band structure, the $s$ and $p$ orbitals of both Ge and S atoms were selected as initial projections. A disentanglement procedure was applied to 16 bands around the Fermi level, which sufficiently span the energy window relevant for the subsequent optical response calculations. Finally, we have prepared the Wannier TB model for shift current calculation.
    \par
    Next, we describe the computational details and parameter settings used for the shift current calculations. We first carefully examined the convergence with respect to $\boldsymbol{k}$-point sampling, as summarized in Fig.~\ref{SM kmesh convergence}. The results demonstrate that convergence is fully achieved when the sampling density exceeds $200\times200$. We then adopted a smearing factor of 30 meV for GeS, treated as a two-dimensional (2D) system. This choice of parameters was consistently applied to all computational approaches.
    \par
    We finally discuss the treatment of band degeneracies. For the sum rule method, an additional broadening parameter $\eta$ must be introduced near degenerate points, for which we used 40 meV. In contrast, for the projector formalism, one has to specify an energy window $\delta\varepsilon$, and the corresponding results for different choices are shown in Fig.~\ref{SM de convergence}. When no special treatment of degeneracies is applied ($\delta\varepsilon = 0$ meV), strong fluctuations appear in certain frequency regions. These fluctuations originate from anomalously large values of projector derivatives near degeneracies. This behavior is also evident in the $\boldsymbol{k}$-resolved results shown in Fig.~\ref{SM kspace plot}, where pronounced spikes (dark-colored spots) emerge if no energy window is applied. By contrast, introducing a finite energy window stabilizes the results, and we find that choices in the range 2$\sim$10 meV yield consistent outcomes, highlighting the robustness of this scheme. Nevertheless, to minimize potential errors, we adopt the smallest possible window, and all results presented in the main text were obtained with $\delta\varepsilon = 2$ meV.
    \input{Figure/SM_GeS_shift_current}

    \input{Figure/SM_kmesh_convergence}
    \input{Figure/SM_de_convergence}

    \input{Figure/SM_kspace_plot}

%% file: Figure/SM_GeS_shift_current.tex
\begin{figure*}[ht]
    \centering
    \includegraphics[scale=0.2]{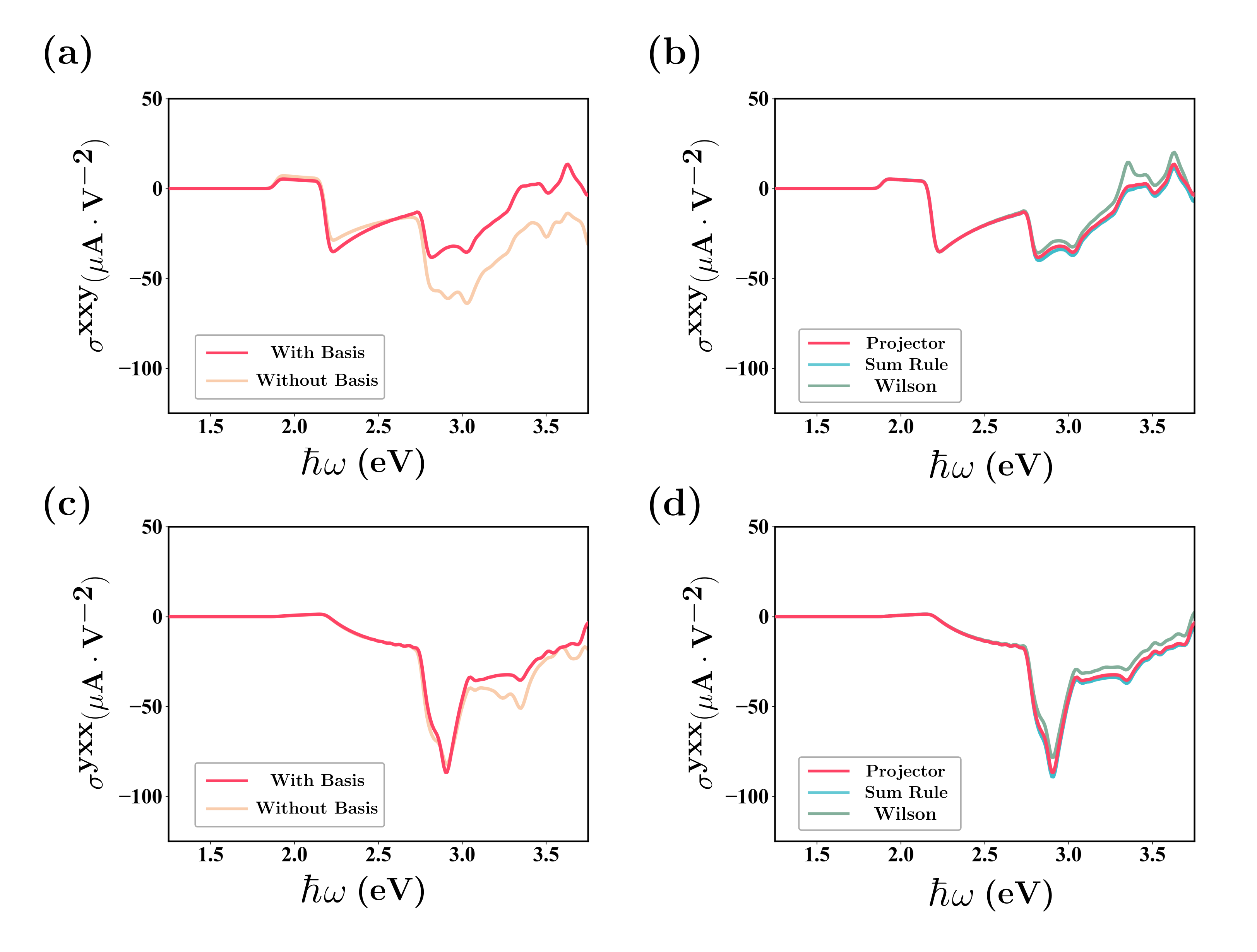}
    \caption{First-principles results of the shift current conductivity $\sigma^{xxy}$ (a~b) and $\sigma^{yxx}$ (c~d) in monolayer GeS. (a, c) Comparison of shift current calculated by the projector method with full basis broadening effect included (red) and neglecting basis broadening (yellow). (b, d) Comparison of shift current calculated by the projector method (red), the sum rule approach (blue), and the generalized Wilson loop method (green).}
    \label{SM GeS shift current}
\end{figure*}

%% file: Figure/SM_kmesh_convergence.tex
\begin{figure*}[ht]
    \centering
    \includegraphics[scale=0.3]{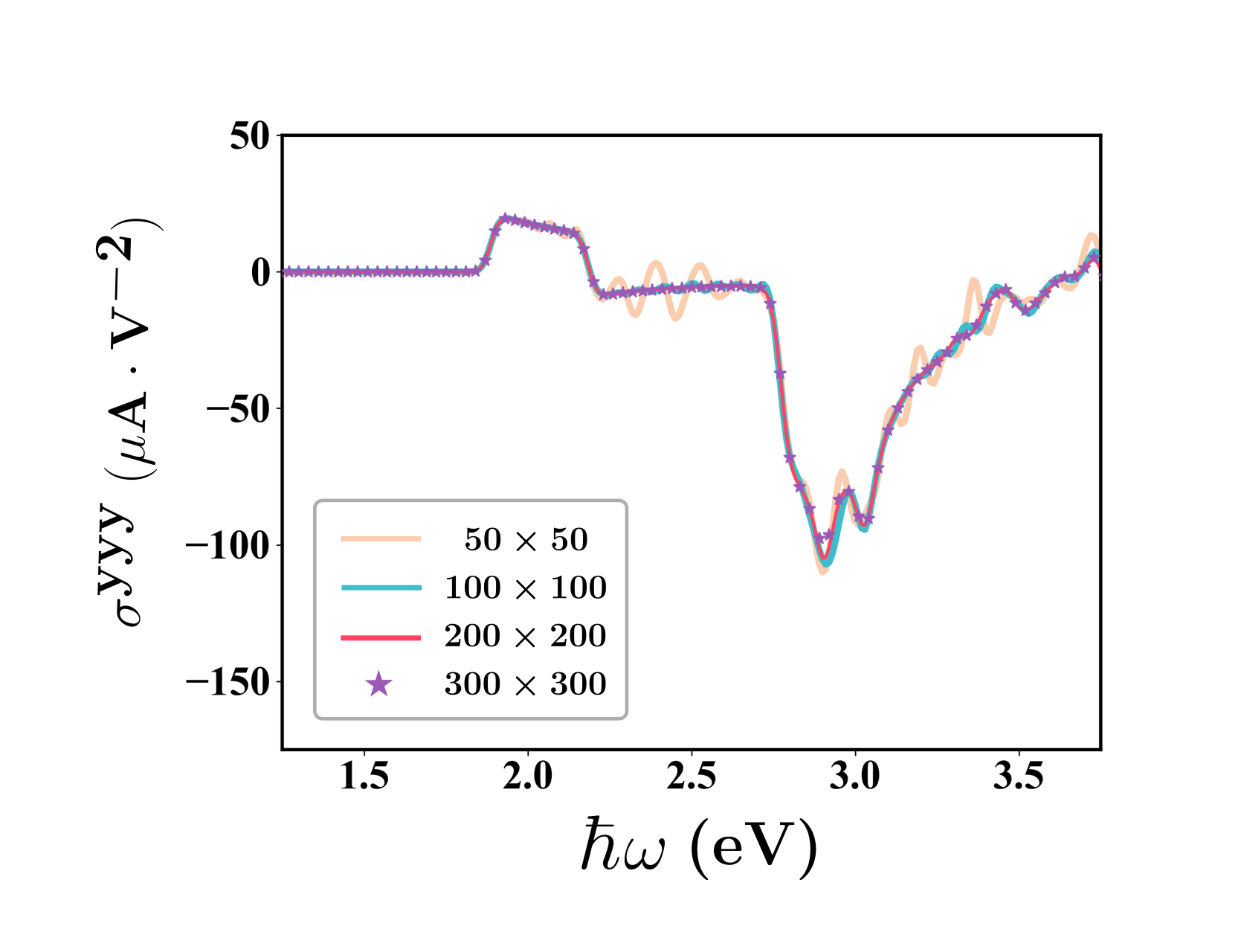}
    \caption{Shift current conductivity with different $\boldsymbol{k}$-point samplings. The results converge for sampling densities of $200\times200$ or higher.}
    \label{SM kmesh convergence}
\end{figure*}

%% file: Figure/SM_de_convergence.tex
\begin{figure*}[ht]
    \centering
    \includegraphics[scale=0.3]{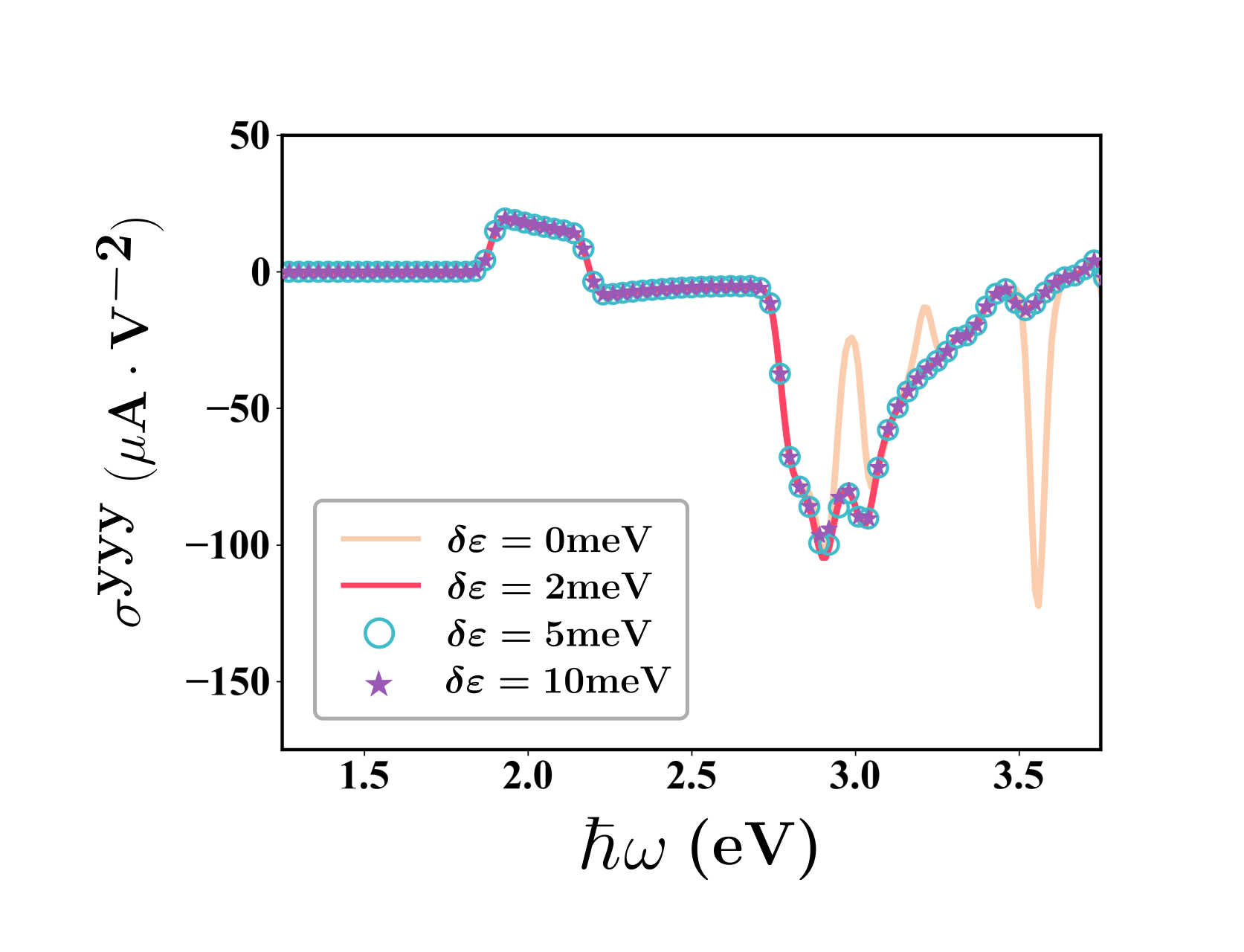}
    \caption{Shift current conductivity with different choices of the energy window $\delta \varepsilon$ for treating band degeneracies. An energy window $\delta \varepsilon$ of 0 meV corresponds to calculations without additional degeneracy treatment. Windows between 2 and 10 meV yield well-converged results.}
    \label{SM de convergence}
\end{figure*}

%% file: Figure/SM_kspace_plot.tex
\begin{figure*}[ht]
    \centering
    \includegraphics[scale=0.2]{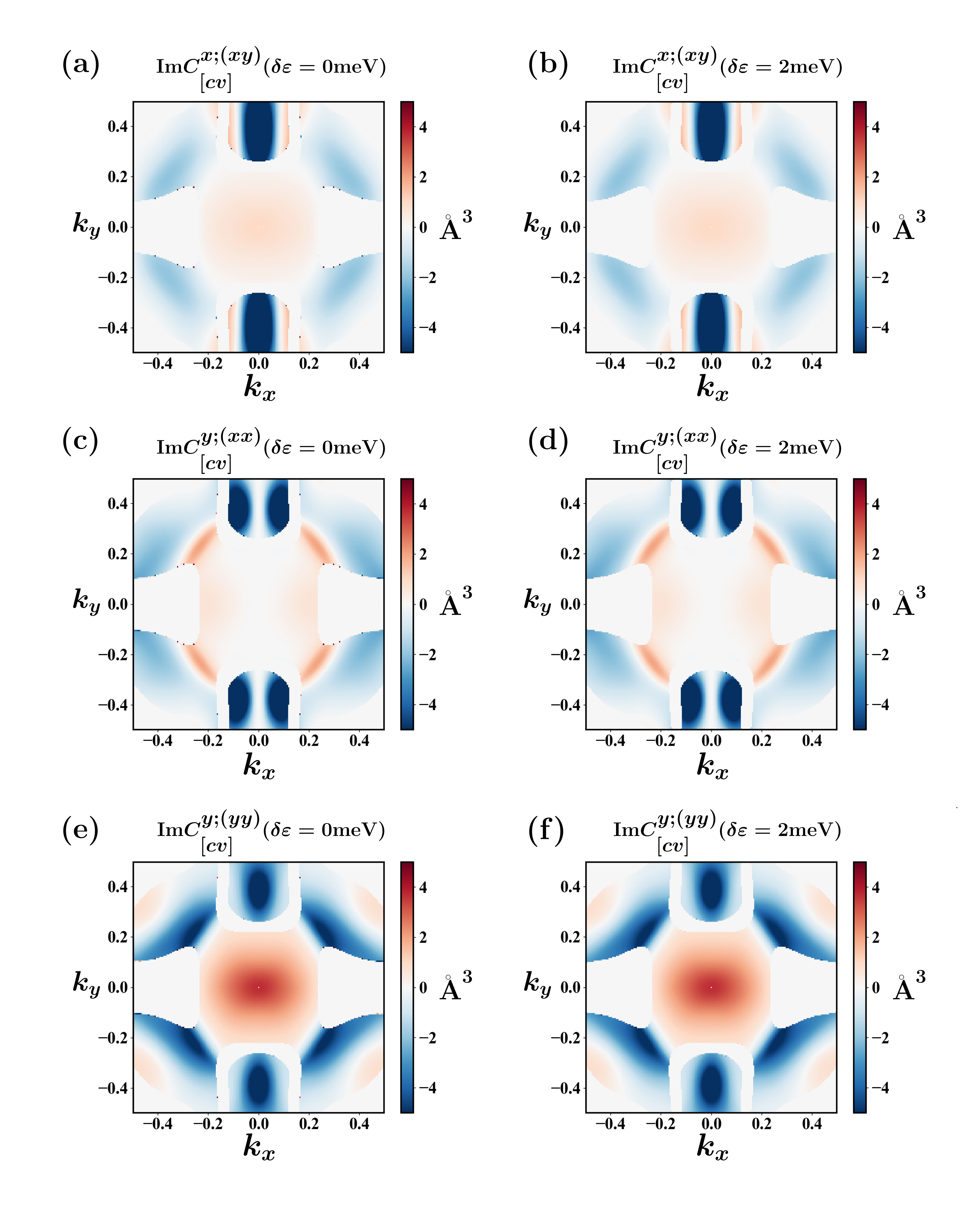}
    \caption{$\boldsymbol{k}$-resolved distribution of geometric quantities Im$C^{\mu; (\alpha\beta)}$ relevant to the shift current. Results are shown for the nearest conduction (c) and valence (v) bands to the Fermi level. $\delta \varepsilon = 2$ meV (0 meV) courresponds to calculations with (without) degeneracy treatment.}
    \label{SM kspace plot}
\end{figure*}